\documentclass{aa}  
\usepackage[dvipsnames]{xcolor}
\usepackage{float}
\usepackage{graphicx}
\usepackage{ulem}
\usepackage{txfonts}
\usepackage[colorlinks=true,linkcolor=magenta,allcolors=magenta]{hyperref}

\begin{document} 

\newcommand{\annot}[1]{\textbf{\color{OrangeRed} #1}}
\newcommand{\reply}[1]{\textbf{\color{Green} #1}}
\newcommand{\beout}[1]{\textbf{\color{Blue} #1}}

\defcitealias{kaastra09}{K9}
\defcitealias{kaastra2011}{K11}
\defcitealias{detmers2011}{D11}
\defcitealias{kaastra2018}{K18}

   \title{A constant pressure model for the warm absorber in Mrk~509}

   \subtitle{}

   \author{Krzysztof Hryniewicz\inst{1}
           \and
           Agata Różańska\inst{2}
          \and
          Tek Prasad Adhikari\inst{3,4}
          \and
          Matteo Guainazzi\inst{5}
           \and
          Biswaraj Palit\inst{2}
          \and 
          Rafał Wojaczyński\inst{2}
          }

   \institute{National Centre for Nuclear Research, Astrophysics Division, Pasteura 7, 02-093 Warsaw, Poland\\
   	\email{krzysztof.hryniewicz@ncbj.gov.pl}
   	\and
   	Nicolaus Copernicus Astronomical Centre (NCAC), Polish Academy of Sciences,
              Bartycka 18, Warsaw, Poland
         \and
         CAS Key Laboratory for Research in Galaxies and Cosmology, Department of Astronomy, University of Science and Technology of China, Hefei, Anhui 230026, China
         \and
         School of Astronomy and Space Science, University of Science and Technology of China, Hefei, Anhui 230026, China
         \and 
         ESTEC/ESA, Keplerlaan 1, 2201 AZ Noordwijk, The Netherlands
             }

  \titlerunning{A CTP model for the WA in \object{Mrk 509}}
  \authorrunning{Hryniewicz et al.}
   \date{Received ...; accepted ...}

  \abstract
     {We present the analysis of 900~ks high-resolution RGS {\it XMM-Newton} observations of the nearby Seyfert galaxy \object{Mrk 509} with the use of a self-consistent warm absorber (WA) model.  We perform a physically motivated approach to the modeling of WA as a stratified medium in constant total pressure (CTP) regime.}
    {Powerful outflows are fundamental ingredients of any active galactic nuclei (AGN)  structure. They may significantly affect the cosmological environment
  of their host galaxy. 
  High-resolution X-ray data are best suited for outflow's studies, and the observed absorption lines on heavy elements are evidence of the physical properties of an absorbing gas.
  Our models allow us to fit continuum shapes bounded together with the line profiles, which gives additional constraints on 
  the gas structure of WA in this source. In this work, we benchmark and test the CTP model on the soft X-ray spectrum of \object{Mrk 509}.}
    {A grid of synthetic absorbed spectra was computed with the photoionization code {\sc titan} assuming that the system was under CTP. As an illuminating spectral energy distribution (SED), we used the most actual multiwavelength observations available for \object{Mrk 509}. We apply these models to the high-resolution spectrum of the WA in the Mrk~509, taking into account cold/warm/hot Galactic absorption on the way to the observer.}
    {CTP gas with $\log \xi_{0} \sim 1.9$, defined on the cloud surface, fits the data well. A higher ionization component is needed for \ion{Ne}{X} absorption. The best-fit model is optically thin with $\log N_{\rm H }= 20.456 \pm 0.016$. The lines are non-saturated, and the CTP spectral fit aligns with previous analyses of \object{Mrk 509} with a constant density WA. The model constrains the gas density, placing the WA cloud at 0.02 pc, consistent with the inner broad line region and the thickening region of the accretion disk.} 
     {}


   \maketitle

\section{Introduction}

Warm absorbers (WA) are clearly seen in about 50\% of active galactic nuclei (AGN)
\citep{reynolds97,blustin2005,laha2014}, and are direct candidates for explaining the feedback of the AGN to the galaxy and the intergalactic medium (IGM)  \citep{tombesi2013,gofford13}. At the very beginning of the gratings detector area on the board of {\it Chandra} and
{\it XMM-Newton}, 
X-ray spectra of ionized outflows in AGN have been studied using the superposition of discrete gas components.
However, it was recognized that this line richness from a variety of heavy elements requires the model resulting from photoionization calculations with some assumption about the matter structure. At that time, the reasonable assumption was that the absorbed gas is of constant density (hereafter CD) \citep[][and many other papers]{kaspi2001,collinge2001,kaastra2002,behar2003,netzer2003,
krongold2003,yaqoob2003,steen2003,blustin2003}.

So far, WAs have been fitted by superposing the effect of CD slabs on different AGN. However, since high-resolution
spectra display lines from different levels of ionization, 
several CD slabs are needed to successfully explain the X-ray spectra \citep{turner2004,steen2005,
costantini2007,winter2010,winter2012,tombesi2013,laha2014}.  
To make clouds of different levels of ionization physically bounded, the more sophisticated 
model was provided by \citet{rozanska2004} assuming that the gas is under
constant total pressure (hereafter CTP). In such a model, gas stratification is provided 
by radiation pressure compression mechanism \citep[see also][]{rozanska2006,rozanska2008,stern2014}.

The direct fitting of the line profile provides us with an estimate of the total energy absorbed by the line, the so-called line equivalent width (hereafter EW). After EWs of observed 
lines are found, and assuming that the observed lines are not saturated, we can derive the ionic column density of each ion responsible for
the given line transition. This ionic column density for non-saturated lines is  proportional 
to its EW, i.e., the linear part of the curve of growth 
\citep[for illustration see:][]{adhikari2019}. 
This kind of work is made directly with the data and is the
first step in the study of WAs. 
The second step is when, after deriving ionic column densities from the data, we can merge them with the total column density of the WA gas by fitting them with modeled ionic column densities computed by photoionization codes.
These total column density, often expressed in terms of 
the hydrogen column density, may be different for each 
CD slab used in the data fitting process.
The third step of the analysis is when we collect them into the so-called absorption measure distribution (hereafter AMD; Holczer et al. 2007; Behar 2009), allowing us to directly measure the strength of the absorption across the AGN outflow. After the AMD of several sources has been determined, it appeared that WAs have the broadest distributions of ionization among all absorbers observed in the Universe. Note that in the second stage of the above procedure, we can use more advanced photoionization models on the stratified density gas and try to reconstruct a continuous WA model, as done by several authors \citep{steen2005,rozanska2006,costantini2007}.

The more sophisticated WA model under CTP was directly fitted to the data only in case of NGC~3783 high-resolution
 {\it Chandra} \citep{goncalves2006}. For the purpose of that paper, photoionization calculations under CTP were made using the {\sc titan} code \citep{dumont2003}.
And because in those days CTP models required large computational power and time, they  have been used to fit ionic column densities computed by {\sc titan} to those derived by observations made by the {\it Chandra} X-ray observatory,  i.e. the second stage of the whole procedure described above. It appeared that for NGC~3783 we cannot distinguish  between 
CD and CTP models based on the relation of ionic column densities to the total column density of WA \citep{netzer2003,goncalves2006}.
 Similar conclusions were drawn by \citet{goosmann2016}, where ionic column densities obtained by radiation pressure compressed cloud modeled with {\sc titan} were nicely fitted to the column densities observed of the particular ion, but the authors did not aim to reconstruct strong deeps in AMD observed for NGC~3783.

On the other hand, CTP models of WA were successfully used to explain the AMD shapes observed in the case
of several AGN \citep{adhikari2015,adhikariamd2019}. 
This work was performed by estimating the AMD shape directly from the computed CTP models, 
without comparing the ionic column densities. The result was that AMD dips clearly observed in WAs are caused by thermal instabilities in the photoionized gas.  Furthermore, the authors presented strong evidence
that this gas has to be quite dense with a density on the order of $10^{12}$~cm$^{-3}$. The exception was Mrk~509, 
where densities of the order of $10^{8}$~cm$^{-3}$ were sufficient to reconstruct AMD in this source \citep{adhikari2015}.
Since absorption models of gas under CTP predict dips in AMD that have been seen in the data
\citep[e.g.][]{behar2009,laha2016}, CTP models should be more suitable to fit the observed lines directly to the X-ray spectra. 
A single CTP model of the gas with density stratification may account for several CD slabs.
However, due to the complexity of model computations (as matter compressed by radiation can be thermally unstable) and spectral fitting (since multiple lines may not fit ideally), this work has never been done.

The ionized outflow in \object{Mrk 509} is the one among several WAs that passed all the steps of the analysis described above. 
The source was observed by high-resolution gratings of {\it XMM-Newton} since 2000. 
The 600 ks observations taken between October and November 2009 were presented by \citet[][hereafter D11]{detmers2011}.
They could successfully fit the high-resolution spectra taken by the Reflection Grating Spectrometer \citep[RGS;][]{denherder01} with a superposition of the components of the CD slab.
After deriving ionic column densities for each of the ion behind the absorption line, they fitted  ionization parameter and hydrogen column density of models which contribute to a given line or ion population.
\citetalias{detmers2011} concluded that the WA in Mrk~509 cannot be described by a smooth continuous AMD, but instead shows two strong discrete peaks. In addition, the authors concluded that there is no single model component that describes all the observed outflows. 
Furthermore, \citet{adhikari2015} confirmed it on the basis of fitting AMD in case of Mrk~509 with CTP model.

Therefore, for the best observed source Mrk~509, there is a discrepancy in our understanding of the WA. Does a contradictory physical description of the nature of the ionized outflow in \object{Mrk 509} emerge from the direct estimate of the AMD, or from the fit of spectroscopic data with discrete CD components.
On the basis of the AMD, the outflow can be continuous, but detailed spectral fitting of high-resolution X-ray data does not support this.
To address this discrepancy, in this paper, we propose to revise the first step of the WA analysis for the source of Mrk~509 i.e. to fit the  RGS spectra, presented in \citetalias{detmers2011}, in the CTP scenario, to understand what is the origin of this 
discrepancy. Does it come from the model limitation or from the lack of our understanding of physical processes in the absorbing gas?   

In this paper, we fit the CTP model directly to the high-resolution X-ray data, taking the information about the gas structure which we have learned from AMD analysis of the same source. Therefore, for the first time, 
the complete WA approach, i.e. estimating the column 
densities with CD components together with AMD derivation, 
is completed with the direct fitting of CTP photoionization model, in case of the one source.
For this purpose, we prepared a grid of CTP {\sc titan} models specifically computed for the \object{Mrk 509} spectral energy distribution (SED) from the multi-wavelength campaign of \citep{kaastra2011a}.
The ionic column densities and the EWs of lines are self consistently derived from the fitted model, and we can check if observed lines are saturated, i.e. on which branch of curve of growth the lines are placed \citep{adhikari2019}.
For this work, we did not use exactly the same data set as 600 ks data used by \citetalias{detmers2011}. 
Alternatively, we stacked together spectra extracted from observations covering the time range between 2000 and 2009
for a total exposure time of 900 ks.

The structure of the paper is as follows: in Sect.~\ref{sec:source}, we present the source and
its WA prior observations. In Sect.~\ref{sec:data},
we describe the data and the reduction procedure used
in this paper. The spectral model used for data fitting is described in Sec.~\ref{sec:model}, while 
the spectral analysis is presented in~\ref{sec:anal}. Section~\ref{sec:resu} presents our results for fitting the total model. These results are discussed in Sect.~\ref{sec:disc} and concluded in Sec.~\ref{sec:conc}.

\section{Warm absorber in Mrk~509}
\label{sec:source}

\begin{figure}
\centering
\includegraphics[width=1.05\hsize]{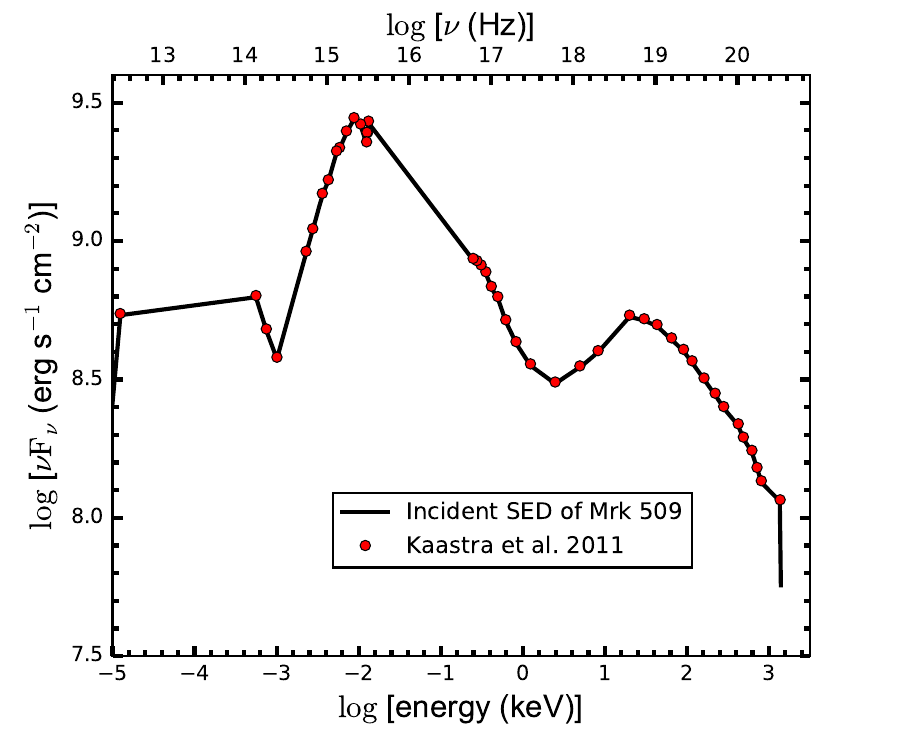}
\caption{Spectral energy distribution of Mrk~509 used as an input to the photoionization code 
{\sc titan} marked as black solid line. The red circles are the observational points reported by \citet{kaastra2011a} and normalised to the incident flux.}
\label{fig:incident509}%
\end{figure}

Mrk~509 is a Sy1.5 galaxy, also considered the closest QSO / Sy1 hybrid, and
it harbors a supermassive black hole of mass
$1.4 \times 10^8$ M$_{\odot}$  \citep{peterson2004}.
It is one of the best studied AGN in the local universe with redshift
of $0.034397$ \citep{huchra1993}
and high luminosity $L(1-1000 ~{\rm Ryd})=3.2 \times 10^{45}$ erg s$^{-1}$, derived from the averaged broad-band SED observed in the multi-wavelength campaign \citep{ebrero2011}.
\citet{kaastra2011a} published a very detailed broadband spectral energy distribution of 
Mrk~509, which was obtained by combining data from their multiwavelength observational campaign using {\it XMM Newton}.  Moreover, their campaign also consisted of simultaneous
data from {\it INTEGRAL} on hard X-rays, as well as {\it Swift}.
The final SED was constructed including the infrared measurements from the {\it IRAS} 
and {\it Spitzer} data and is presented in Fig.~\ref{fig:incident509} (red circles).

Earlier work on outflow in the X-ray regime has revealed that it consists of a wide range of ionization components but lacks a very high and also a very low ionized gas \citep[weak Fe UTA and no \ion{Si}{XIV} Ly$\alpha$ as reported by:][]{yaqoob2003}. Usually, three outflow velocity components are needed to fully explain the observed absorption spectra
\citep[][detmers2011]{smith2007}. Later, the X-ray outflow in this source has been extensively studied using the $600$ ks
RGS on board of the {\it XMM-Newton} X-ray telescope
\citep{kaastra2011a,kaastra2012}. The high spectral resolution of RGS allowed one to identify a few tens of absorption lines from highly ionized metals
and hence to determine their ionic column densities \citepalias{detmers2011}. Furthermore, multiple absorption lines from the
interstellar medium were detected. Finally, to fully explain the RGS spectrum, several emission components were needed. 

The above data allowed us to put new constraints on WA in Mrk~509. \citetalias{detmers2011} found that the ionized absorber consists of two velocity components  $v=-13 \pm 11$~km~s$^{-1}$ and $v=-319 \pm 14$~km~s$^{-1}$
which are consistent with previous results including UV data.
Another tentative component $v=-770 \pm 109 $~km~s$^{-1}$  is only seen in a few highly ionized absorption lines. However, several ionization components were assigned to each of the three velocity components. 
Thus, from the derived ionic column densities and with the use of photoionization calculations, \citetalias{detmers2011}
determined the equivalent hydrogen column densities and constructed the AMD.
The authors demonstrated that the outflow in Mrk~509, in the $\log(\xi)$ range between 1 and 3 can be described by a broad function with two pronounced dips, where the column density
drops by 2-3 orders of magnitude. Later, \citet{adhikari2015} confirmed that such deeps are caused by thermal instability in the structure of the CTP cloud.

In addition, \citetalias{detmers2011} showed that the highest ionization component is located at 0.5 pc,
assuming that the observed changes in the absorption spectrum
are due to the response to changes in the ionizing continuum with a delay related
to the electron density. 
\citet{kaastra2012} estimated the upper limits on the location of different ionization
components of the warm absorber in the range of $5$ to $400$ pc. All photoionization calculations performed by these authors were performed with the assumption of CD and the input SED determined for Mrk~509 from observations (Fig.~\ref{fig:incident509}).

\section{Data reduction }
\label{sec:data}

We extracted Observation Data Files (ODF) for all the observations of Mrk~509 available in the public XMM-Newton archive\footnote{Obs.\#0130720101, \#0130720401, \#0306090101, \#0306090201, \#0306090301, \#0306090401, \#0601390201, \#0601390301, \#0601390401, \#0601390501, \#0601390701, \#0601390801, \#0601390901, \#0601391001, \#0601391101}. They correspond to three separate campaigns performed in October 2001-April 2001, October 2005-April 2006, and October-November 2009. Data for each ODF were reduced using SASv16.0 \citep{gabriel04}, using the most updated calibration files available at the time of data reduction (October 2019). Calibrated event lists, source and background spectra and response files (combination of the redistribution matrix and of the energy-dependent effective area) were generated with the SAS reduction meta-task {\tt rgsproc}. The reference position for the energy scale reconstruction was assumed at the optical coordinates of the AGN. {\tt rgsproc} was run twice. The event lists generated in the first run were used to extract light curves of the CCD number 9 from a spatial region outside the dispersed source spectrum. In each light curve, intervals with a high particle background were identified as those corresponding to a count rate $>$2~s$^{-1}$. In a second {\tt rgsproc} run, the time intervals corresponding to the low particle background were used as input to select good events, from which the final scientific products were extracted.

Following \citetalias{detmers2011}, we combined the RGS\,1 and RGS\,2 spectra and responses generated in each observation with the SAS task {\tt rgscombine}. The total exposure times are 903 and 900~ks for RGS\,1 and RGS\,2, respectively. In their analysis of the deep RGS observational campaign of Mrk~509, \cite{kaastra2011a} and \citetalias{detmers2011} used the combined fluxed spectrum instead of the combined count spectrum, to avoid spurious narrow-band features that could arise at the wavelength of individual unexposed pixels because {\tt rgscombine} combines the instrumental response without weighting \citep{kaastra2011a}. This issue could be exacerbated by the fact that the RGS was operated in ``multi-pointing mode'' during the 2009 observational campaign. However, fitting the fluxed spectrum introduces systematic uncertainties that are difficult to quantify because of the approximation required to correct the observed count spectrum for the instrumental response. The range of Mrk~509 count rates measured by the RGS is moderate (0.86--1.50~s$^{-1}$ in the RGS\,1 0.2--2.0~keV energy band, for example) except in the comparatively short 32~ks, Obs.\#0130720101 which registered a RGS\,1 count rate of 0.55~s$^{-1}$ in the same energy band. We estimate that the systematic uncertainties potentially induced by the combination of the effective areas without weighting is $\le$1.5\%, that is, comparable to the systematic uncertainties in the calibration of the effective area \citep{devries15}. In light of this result, we prefer employing a rigorous forward-folding approach on the combined count spectra. 

The above procedure allowed us to obtain a signal-to-noise ratio of the order of 10 on average, which is three times less than the signal-to-noise ratio reported by \citetalias{detmers2011}, after combining both RGS detectors. In order to prevent any additional lowness of the signal-to-noise ratio, we regret to make any data binning, even if the {\tt ftgrouppha} tool from Heasoft ver. 6.25, with the option of optimal, indicates the
saturation limit in the binning of high-resolution data \citep{kaastra2016}.   
For visualization of the overall shape of the spectrum, we draw the data binned by 5 channels in Fig.~\ref{fig:cont}.

\section{Model description}
\label{sec:model}

We consider models of absorbing gas where radiation interacts with matter through photoionization processes. In this setup, we assume an ionizing continuum that illuminates a slab of gas in plane-parallel geometry.  The gas is parameterized by the density of local numbers on the face of the cloud, $n_{\rm H,0}$, and the total density of the columns, $N_{\rm H}$. The total pressure, which is the sum of the gas and radiation pressures,
is assumed to be constant \citep[CTP;][]{rozanska2006}. Our model differs from the radiation pressure compression
model \citep{stern2014,bianchi19}, since in the latter model, only the first order of radiation pressure decreasing proportionally to ${\rm e}^{-\tau}$  is taken into account, while the
{\sc titan} code computes the local value of the radiation pressure from the intensity field
taking into account the current value of a source function \citep{adhikari2019}.

The ionization parameter $\xi_0$, in erg~cm~s$^{-1}$, is related to the ionizing flux by the following expression
\begin{equation}
\xi_0=\frac{L}{n_{\rm H,0} r^2},
\end{equation}
where $L$ is the incident luminosity integrated over the energy range available from observations \citep{kaastra2011a},
and $r$ is the distance from the ionizing source. In the constant pressure regime, the values $n_{\rm H}$ and $\xi$ are strongly stratified across the inner layers of the slab, and we define them only at the cloud surface \citep{rozanska2004,rozanska2006}. 
This is a static solution and does not account for the time evolution or the flow of matter.

A complete grid of models of the absorption spectra for WA was calculated with {\sc titan} code~\citep{dumont2003, chevallier2006}. 
The code solves the transfer of radiation assuming
the Accelerated Lambda Iteration (ALI) method, which takes into account the
full treatment of the source function at each depth of the cloud and
computes the line and continuum intensity self-consistently. The detailed
description and advantages of the ALI method are described in \citet{dumont2003} in the context of the X-ray absorber.  The code computes self-consistently full profiles for absorption and emission lines and takes these profiles into account when doing radiative transfer computations. {\sc titan} can also account for the effects of turbulence by setting the turbulent velocity value.  The CTP medium for the purpose of Mrk~509 was illuminated with SED determined for this source from observations. 
The input SED to the {\sc titan} code is taken from \cite{kaastra2011} (solid black line in Fig.~\ref{fig:incident509}).

The computed grid of ionizing parameters spans a range of 0 to 5 in $\log \xi_0$ with a 0.1 step and 20 to 23 in $\log N_{\rm H}$ with a 0.1 step.
The upper limit of the column density was chosen on the basis of the shape of the absorbed continuum. For $\log N_{\rm H} > 23 $ the overall absorption of the continuum is high enough that it is inconsistent with the observed RGS continuum. 
 We computed the table models for three initial volume densities on the illuminated face of the cloud: $n_{\rm H,0}=10^8, 10^{10}$ and $10^{12}$~cm$^{-3}$.
 The choice of $n_{\rm H,0} \geq 10^8$ cm$^{-3}$ for the \object{Mrk 509} absorber is motivated by the work of \citet{adhikari2015},
 where they demonstrated that the observed AMD for this source can be well reconstructed
with the {\sc titan} models for this gas density. In addition, the general shape of AMD for several sources is
well reconstructed by high-density outflow \citep{adhikariamd2019}.

The {\sc titan} code fully takes into account the influence of turbulent velocity on the modeled line profile.
The effect is illustrated in Fig.~\ref{fig:turb}. It is clearly visible that turbulent velocity modifies the absorption line profile, as expected from the theory
of radiative transfer and the curve of growth analysis. Therefore, we have calculated all sets of models including turbulent velocities relevant
to the values generally observed in other data sets, i.e. $v_{\rm turb} = 100, 300, 500$ and 
$700$~km~s$^{-1}$. Note that turbulent velocities do not influence the shift of the location of the line centroids, which is caused by the outflow velocity.
We did not fit the line shift individually; instead, the bulk shift was fitted by using redshift for the absorption spectra as the independent parameter.

The number of ionic transitions taken into account is 4141 in the energy range between 0.01 eV and 25 keV. The ten most abundant elements are taken into account: \element{H}, \element{He}, \element{C}, \element{N}, \element{O}, \element{Ne}, \element{Mg}, \element{Si}, \element{S} and \element{Fe} with {\it solar} abundances following \cite{grevesse89}.

\section{Spectral analysis}
\label{sec:anal} 

To analyze X-ray spectra, we use the software package Bayesian X-ray Analysis (BXA) \citep{buchner2014},
which connects the nested sampling algorithm {\sc MultiNest} \citep{feroz2008,feroz2009,feroz2019}
with the fitting environment {\sc XSPEC} ver. 12.10.1 \citep{arnaud1996}. 
The procedure was performed in the energy range 0.326--1.550 keV (8--38 \AA).
In the process of evaluating the accuracy of the model, we used the goodness of fit statistic
C-stat \citep{cash1979}, as it is implemented in {\sc XSPEC} \citep{kaastra2017}.
RGS \,1 and RGS \,2 were simultaneously fitted, allowing only a cross-normalization 
factor to remain free in the fit \citep{devries15}. 
We use uniform priors, so no preferences were given over the assumed parameter space.

In the first step, we fitted the model of ionizing the SED without warm absorber lines to have an initial point of reference for further improvements. The continuum model multiplied by Galactic absorption we 
used can be described as a product of components:
\begin{equation}
F(E) = \Pi_j e^{-N_{\rm H,Gal_{\it j}} \sigma} \times A \; S(E),
\end{equation}
where $j$ is a number of Galactic absorption components,
$N_{\rm H,Gal_{\it j}}$ is the  column density of neutral hydrogen in the Milky Way, $\sigma$ is the cross section for absorption and $A$ is the normalization of the continuum. $S(E)$ is the function that describes continuum radiation, which in our analysis is the SED of \object{Mrk 509} that we used as the ionizing continuum in
our photoionization computations. 

\begin{figure*}
   \centering
   \includegraphics[width=0.90\linewidth]{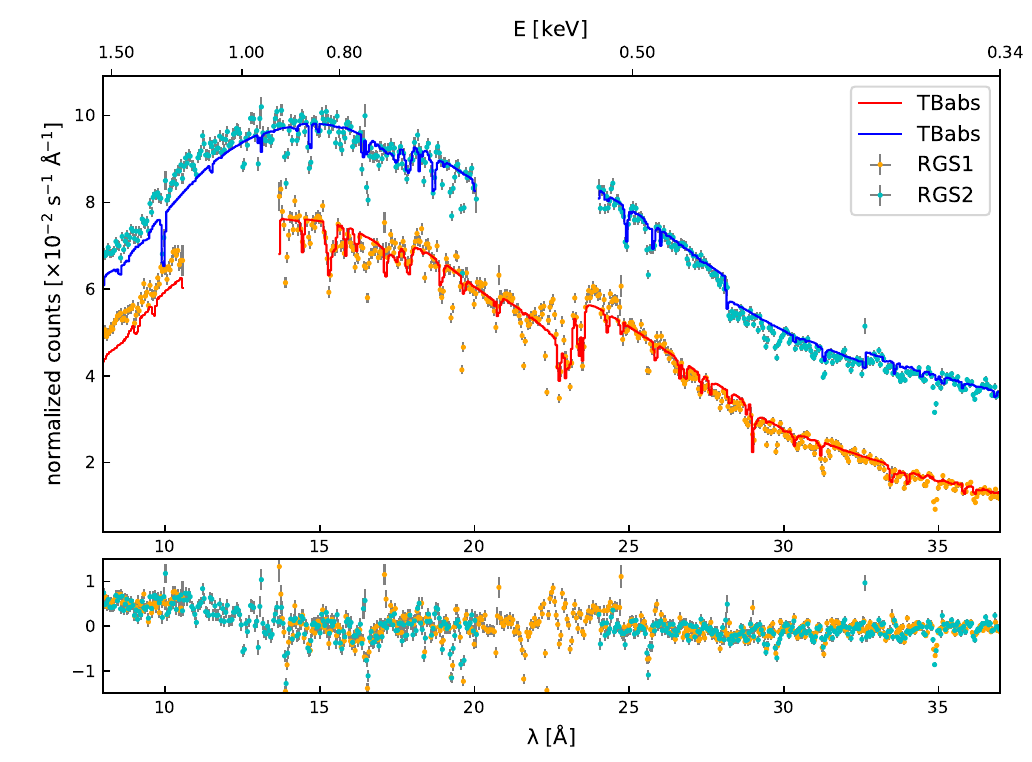}
   \caption{Continuum model and Galactic absorption
   fitted to \object{Mrk 509} RGS spectrum. RGS\,1 and RGS\,2 spectra are plotted in orange and cyan. RG\,2 is arbitrarily shifted by 0.02 counts s$^{-1}$ \AA$^{-1}$.
     The continuum model of \object{Mrk 509} SED absorbed by the Milky Way is marked with red and blue continuous lines. The lower panel shows residuals against the best-fit model. The wavelengths are in the observed frame of reference. For the purpose of this plot the data are binned by 5 channels.}
              \label{fig:cont}
  \end{figure*}

\begin{figure}
\centering
   \includegraphics[width=0.99\linewidth]{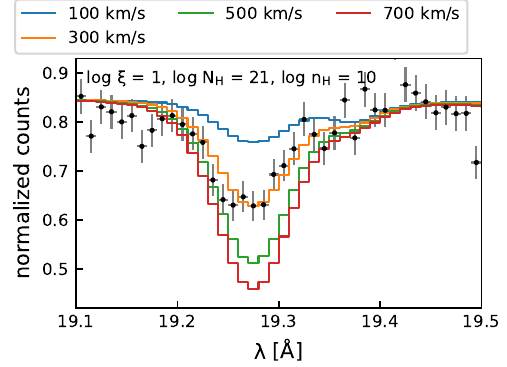}
   \caption{The dependence of O{\sc vii} profile on the turbulent velocities used in  photoionization 
   computations by {\sc titan} code for a given model.}
              \label{fig:turb} 
\end{figure}

 As typical for AGN, the complex absorption can be divided on external absorption connected to our Galaxy, for which the line redshift is almost zero, and intrinsic absorption is usually represented as a WA component, where the lines have the source redshift. As the first-order approximation of Galactic absorption we use a single {\sc tbabs} model with a fixed hydrogen column density value of $4.44 \times 10^{20}$ cm$^{-2}$ \citep{murphy1996}. The overall model of the continuum absorbed by the Milky Way, fitted to the data,
is presented in Fig.~\ref{fig:cont}. The model already displays several absorption lines
due to the Milky Way gas. 

To account for more complex, most likely multiphase Galactic absorption \citep{pinto2012}, we also fit more advanced models: {\sc ismabs} \citep{gatuzz2015}
and {\sc ioneq} \citep{gatuzz2018}. The results of this fitting procedure are presented in the
Appendix~\ref{app:lines}. The above analysis allowed us to
identify several Galactic absorption lines in the data, given in Tab.\ref{tab:ism_lines}, which are later taken into account while identifying WA lines in Sect.~\ref{sec:walines}.  However, for further investigation of the WA components, we postpone the components {\sc ismabs} and {\sc ioneq} and rely exclusively on the model {\sc tbabs} for Galactic absorption. This decision was motivated by prohibitive computing time in {\sc XSPEC} with the Bayesian approach and, in some cases, failures in achieving the fit. 

 In addition, care was taken to identify instrumental features that may not be included in the response matrix of the RGS. The instrumental lines, together with the references, are listed in the Appendix~\ref{app:lines}, Tab. ~\ref{tab:instrument_lines}.

  \begin{figure*}
   \centering
   \includegraphics[width=0.99\linewidth]{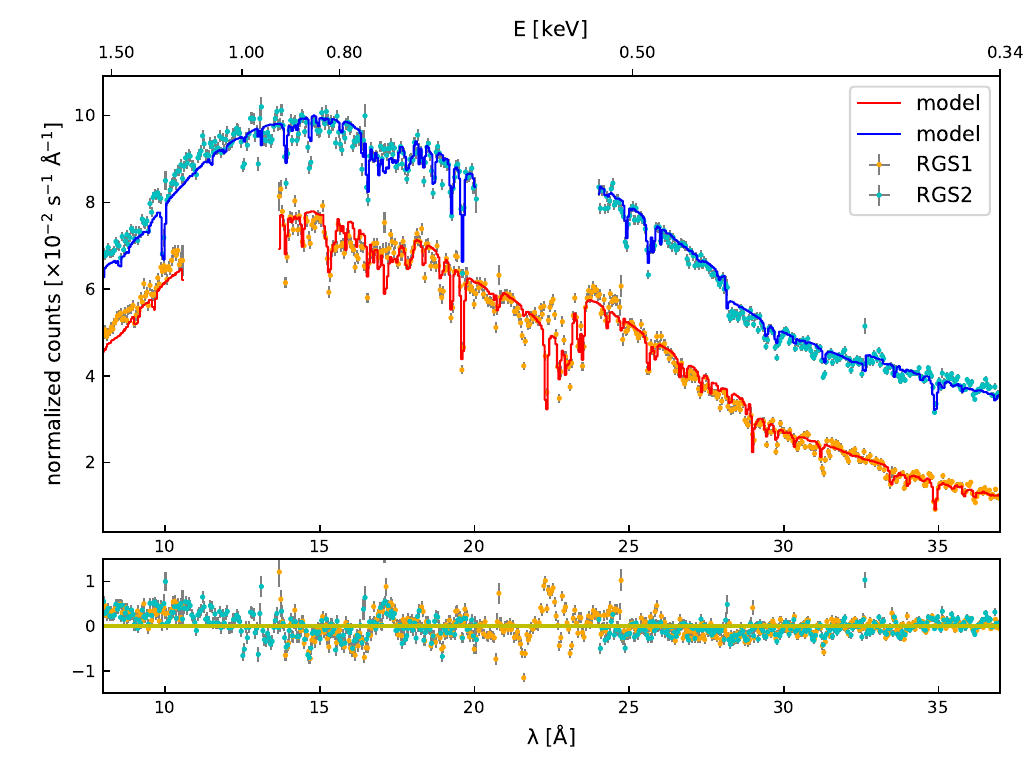}
   \caption{Warm absorber on top of continuum modified by Galactic absorption with \object{Mrk 509} RGS data points. RGS\,1 and RGS\,2 spectra are plotted in orange and cyan. RGS\,2 is arbitrarily shifted by 0.02 counts s$^{-1}$ \AA$^{-1}$. The total model: continuum together with a single CTP WA is plotted in red and blue. The wavelengths are in the observed frame of reference. The lower panel shows residua. For the purpose of this plot the data are binned by 5 channels.}
              \label{fig:twa}
  \end{figure*}

\begin{figure*}
\includegraphics[width=0.99\linewidth]{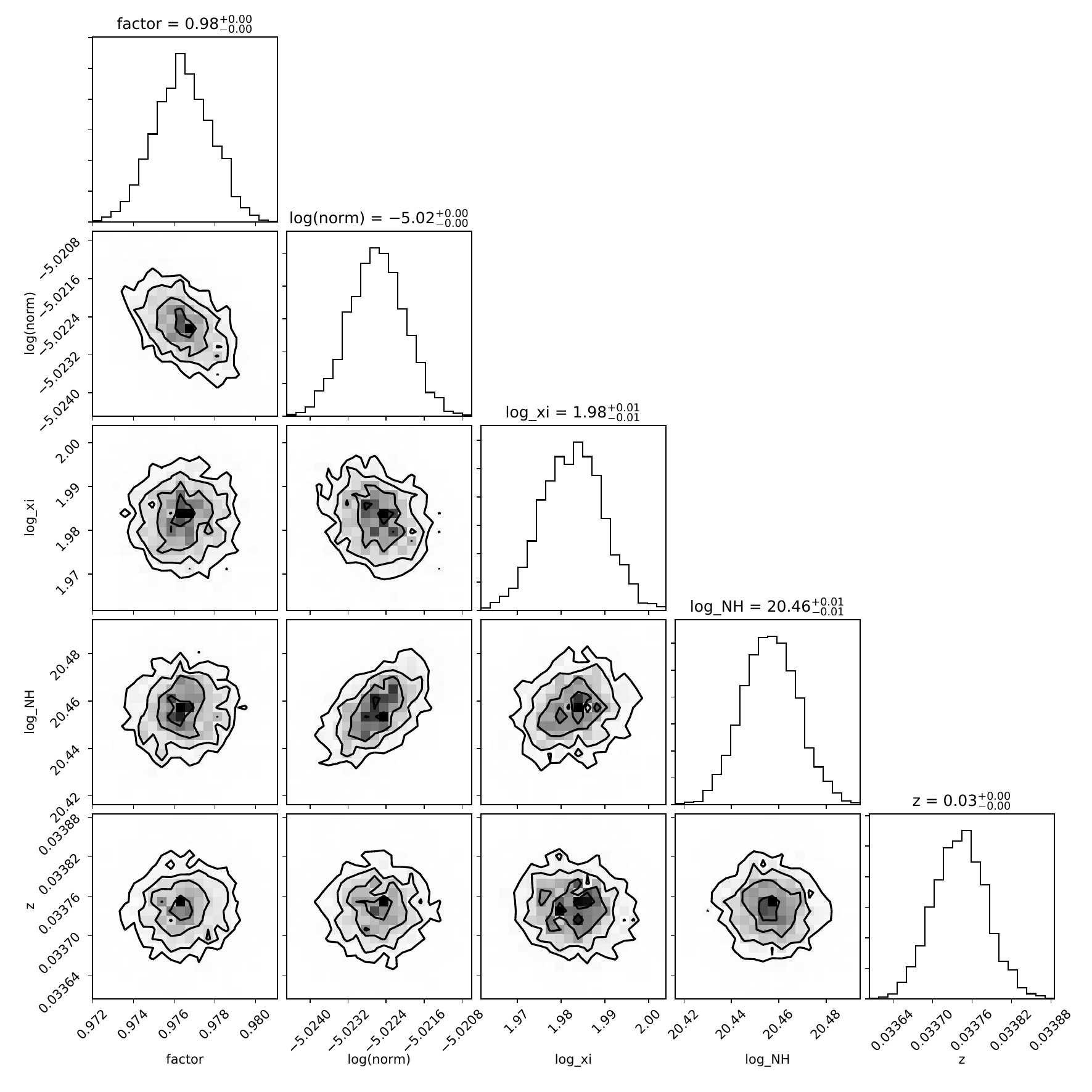}
   \caption{BXA contour plots for fitted parameters of the total model to the \object{Mrk 509} RGS\,1 and RGS\,2 spectra. Cross normalization factor shows difference between two RGS specra where normalization gives the position of the total model. Best fit  log $\xi$ = 1.9829 $^{+0.0093}_{-0.0089}$ and log $N_{\rm H}$ = 20.456 $\pm$ 0.016 with z = 0.03374 $\pm$ 0.00005 corresponds to the WA outflowing with $v=195\pm16$~km~s$^{-1}$. The diagonal panels show marginalized posterior probability distributions while off-diagonal panels indicate conditional probability distribution functions among each parameter.}
   \label{fig:contour_cfadd}
\end{figure*}

\begin{figure}
\centering
   \includegraphics[width=\linewidth]{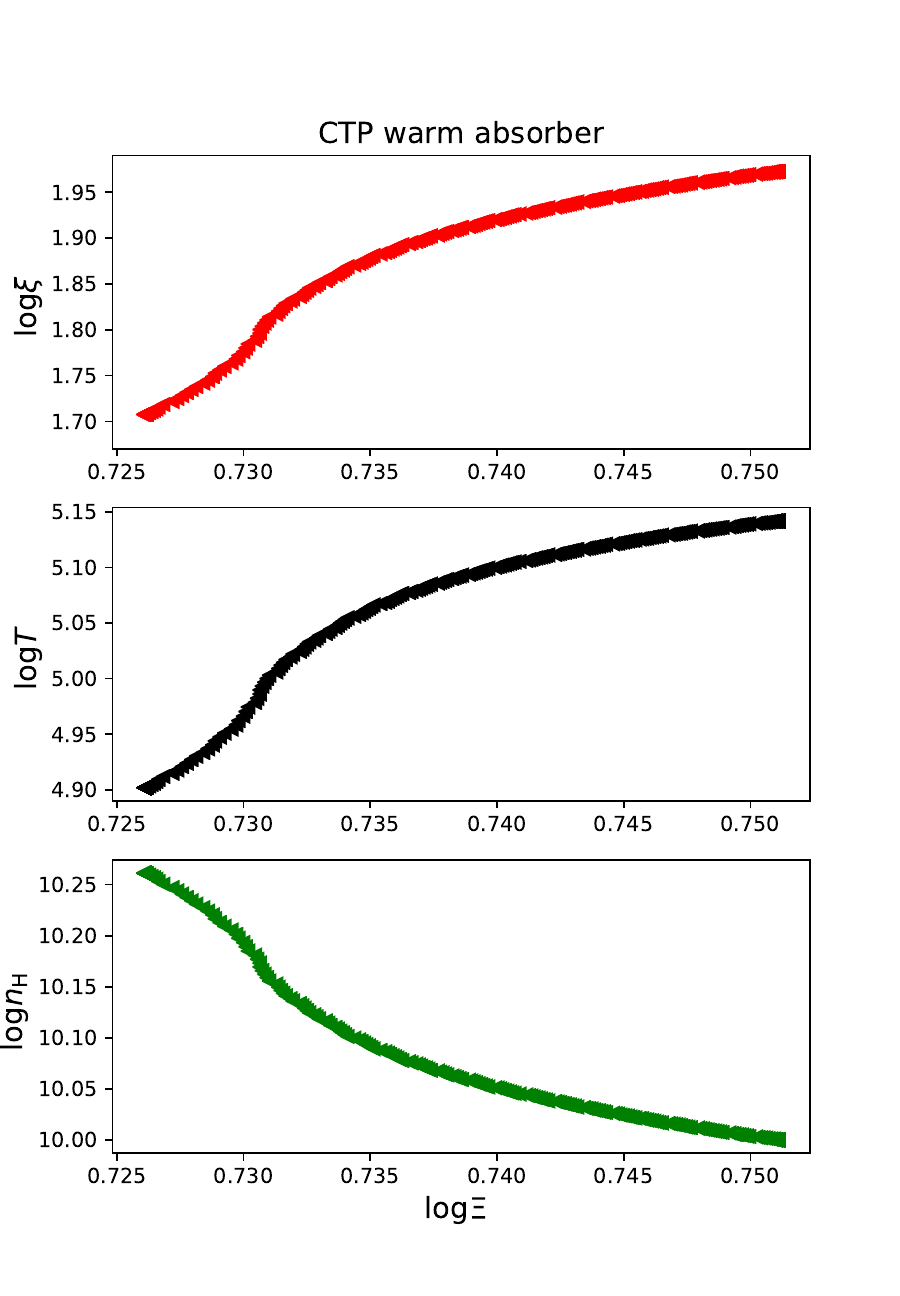}
   \caption{The ionization structure of the best fit CTP model
   computed by {\sc titan} and fitted to the Mrk~509 RGS data. The upper panel shows the structure of ionization parameter $\xi$, medium panel - temperature $T$ (S-curve), and bottom panel - volume density 
   $n_{\rm H}$, all versus dynamical ionization parameter $\Xi$. The illuminated cloud surface is located on the right side of panels.}
              \label{fig:ionization} 
\end{figure}

\begin{table*}
\caption{The best fit parameters of the final model. }
\label{tab:stat}
\begin{center}
\begin{tabular}{llcccccc}
\hline\noalign{\smallskip}
component & $\log N_{\rm H}$  &  $\log \xi_0$ & $\log n_{\rm H,0}$ &  $v_{\rm turb}$ & $\log A  $ & $v$ & c-stat/dof\\
   & [cm$^{-2}$] &  [erg~cm~s$^{-1}$] &[cm$^{-3}$] & [km~s$^{-1}$] & normalization & [km~s$^{-1}$] & \\
\noalign{\smallskip}\hline\noalign{\smallskip}
  SED Mrk509  & --  & --& --& --& $-5.02272 \pm 0.00079 $& -- & -- \\
 {\sc tbabs}   & 20.65  & --& --& -- & --  & -- & 14360/5895 \\
 {\sc titan} (CTP) & 20.456 $\pm$ 0.016 & 1.9828$_{-0.0089}^{+0.0093}$ & 10 & 500 & -- & -195 $\pm$ 16 & 10972/5893 \\ 

\noalign{\smallskip}\hline
\end{tabular}
\end{center}
\tablefoot{
The model is composed of three multiplicative components: continuum multiplied by Galactic absorption {\sc tbabs} and by single warm absorber CTP model, according to the Eq.\ref{eq:tot}. The intrinsic continuum is in the shape of SED given in Fig.~\ref{fig:incident509}. The parameter given without errors were frozen, and final analysis was made on 5893 degrees of freedom.
The additional free parameter was the multiplicative factor which
account for the normalisation difference between RGS\,1 and RGS\,2:
$0.9765_{-0.0019}^{+0.0018}$. The velocity given in column 7 is the kinematic shift in the source rest-frame.
}
\end{table*}

Subsequently, on the top of the continuum modified by Galactic absorption, we added ionized CTP components to the product of models:
\begin{equation}
F(E) =  \Pi_j e^{-N_{\rm H,Gal_{\it j}} \sigma} \times A \; S(E) \times \Pi_i WA_i(N_{{\rm H},i}, \xi_i),
\label{eq:tot}
\end{equation} 
where $i$ is a number of components of the warm absorber and each
$WA_i(N_{{\rm H},i},\xi_i)$
is parameterized with the column density and the ionization parameter.
To be in agreement with AMD fitting, only one CTP component should be needed
to fit the data, that is, $i=1$.

The overall, major fitted components are listed in Tab.~\ref{tab:stat}, where in the third row, we describe the best-fitted parameters of a single CTP model with solar abundances. 
The model requires a turbulent velocity equal to 500~km~s$^{-1}$.

\section{The results of fitting the total model}
\label{sec:resu}

In the following, we discuss our final fit model with its physical implications. 
The overall spectral fit for both RGS data, which includes continuum emission multiplied by Galactic absorption, and multiplied by the single CTP warm absorber model, is presented in Fig.~\ref{fig:twa}. The best-fit parameters are described in Tab.~\ref{tab:stat}.

We point out here that our continuum emission has the shape of an observed SED and was not corrected for unknown intrinsic absorption. In the work of \citet{detmers2011}, the same continuum was used, but the lines were fitted separately as absorption of the ion population. The approach of the authors described above is similar to using absorption lines on top of the incident continuum. 
Our model computations allow using the so-called {\it outward} spectrum, which is the result of the incident spectrum corrected by absorption. 

Thus, in our approach, the best-fitted model for the continuum is a multiplication of the CTP warm absorber model by the incident continuum SED with 100\% coverage, and it is presented in Fig.~\ref{fig:twa}. 
More consistent additive models, where the sum of the incident and outward models (and optionally reflection) is fitted, give an overall good fit  but do not fit so well the \ion{O}{VIII} $\lambda \sim$ 19.6\AA\ and \ion{O}{VII} $\lambda \sim$ 22.33\AA\ lines in the global solution. Weaker lines in the global fit of the additive solution are probably caused by the fact that the incident component is stronger than the outward component to preserve a good overall fit of the continuum. This is interpreted as a lower covering factor of the absorber. But this destroys fit in the deeper lines profiles. 
Thus, we conclude that the true incident continuum should be softer with a higher level of flux density above 25\AA\ (below 0.5 keV).
Having this in mind, we decided to stay with the original input SED, even if our fit is not perfect. 
However, since we do not have any better constraints on the shape of the continuum, we have chosen to use the original SED, which allows the comparison of our results with the previous work.

The Galactic column density of our overall fitted model with component {\sc tbabs} is 
$N_{\rm H}=4.44  \times 10^{20}$\,cm$^{-2}$, and it is kept frozen during spectral analysis 
\citep{murphy1996,Willingale2013MNRAS}. 
It should be noted that the above value is 0.5 higher than that found by HI4PI \citet{2016HI4PI}.

We made the initial choice of the best $n_{\rm H,0}$ by comparing independent fits of the models and assessing their fit quality.
For each of $n_{\rm H,0}=10^8, 10^{10}$ and $10^{12}$~cm$^{-3}$ we fitted the overall absorption spectrum in the parameter space $\log \xi_0$--$\log N_{\rm H}$. The best statistic was obtained for $n_{\rm H,0}=10^{10}$~cm$^{-3}$ although the advantage over $10^{8}$~ cm$^{-3}$ was rather minor on the order of a few hundred in the c-stat value. 
This consideration is also in line with the conclusion that AMD analysis does not exclude models with gas density $10^{10}$ cm$^{-3}$ \citep[see Fig. 14,][]{adhikariamd2019}.
Thus, we further proceed our analysis with the model assuming $n_{\rm H,0}=10^{10}$~cm$^{-3}$.

In a similar fashion, we made a choice of the turbulent velocity in the model.
From the calculated grids with $v_{\rm turb} = 100, 300, 500$ and $700$~km~s$^{-1}$ we chose the velocity for which the best value of c-stat was obtained. Effectively $v_{\rm turb}$ increases the depth of absorption for the given $\xi$, $N_{H}$ and $n_{H,0}$ as illustrated in Fig.~\ref{fig:turb}. The best result was achieved for $v_{\rm turb} = 500$~km~s$^{-1}$.

The parameters of the best-fit
WA model are: $\log \xi_{0} = 1.9828_{-0.0089}^{+0.0093}$,
$\log N_{\rm H} = 20.456 \pm 0.016$.
The fitted values with their errors are interpolated values as given by the BXA package, while the grid step was 0.1 in both parameters, and this value defines a potentially more realistic precision of the solution.
The overall fit statistic in our analysis is worse than the one obtained by \citetalias{detmers2011}, but a direct comparison of the statistic is not possible. On the one hand, we did not add both RGS spectra, allowing for the lower statistic, but on the other hand, we did not bin the data in order to keep all lines visible even on the detection limit. In addition, our model is also more tight, since the entire photoionization component in the CTP regime fits all lines for a given ionization parameter on the
cloud surface \citep{adhikariamd2019}. Alternatively, \citetalias{detmers2011} needed 5-6 constant density components to model lines from different ionization states of heavy elements, and such approach is much more flexible 
in terms of goodness of fit. 

 The ionization parameter on the cloud surface, $\xi_0$, resulting from our fitting is relatively low $\log \xi = 1.98$, and decreases as the depth of the WA cloud increases, as presented in the upper panel of
Fig.~\ref{fig:ionization}. The temperature and density structure given in the lower panels
directly follow this change. The illuminated cloud surface is on the right side of Fig.~\ref{fig:ionization}. Even if ionization is stratified in our model, it does not reflect the high-ionization constant density components fitted by \citetalias{detmers2011}.  Our WA of best fit is optically thin with total column density $\log N_{\rm H} = 20.456$. 
The first reason for this fact is that, in a CTP regime, ionization fronts are very narrow, and with only a minor change in temperature and density, we jump to a different ionization state when lines appear. It is worth pointing out that the resulting values of the temperature of the stratified WA are outside the regime where typical thermal instability occurs.

The second reason is that the collected RGS data cover a very narrow energy range,
i.e. 0.34-1.5~keV, and the data cannot fully reflect the highly ionized CTP phase.
For the same reason, the broad distribution of ionization, which manifests itself in
absorption measure distribution constructed with the use of constant density slabs, can be modeled only by a
relatively narrow optically thin CTP model. 
A new high-resolution
detector with energy coverage up to 10~keV, as the X-IFU instrument \citep{2023Barret} on the board of the future New-Athena mission, is needed
to account for the optically thick WA consisting of low- and high-ionization zones
in the single CTP model \citep{rozanska2008}.

Fig.~\ref{fig:contour_cfadd} shows the confidence contours generated by BXA between different parameters of the WA CTP model that best fits. 
The cross normalization factor between the RGS \,1 and RGS \,2 spectra is very well constrained to a value of 0.98. A small degeneracy between normalization and total column density is observed, but otherwise parameter values are robustly constrained. This suggests that our fit is acceptable for further analysis.

For many years, the dependence of photoionization models on the gas volume density was thought to be degenerate, i.e. the absorbed spectra were the same for wide ranges of the gas density.
But this conclusion was valid for early studies of optical absorption lines from nebulae, where the physical conditions of matter required low-density gas, of the order of $\sim 10^4$\,cm$^{-3}$.
Due to this reason, constant density models were generally accepted in publicly available photoionization codes. But with increasing sensitivity of spectrometers it appeared that this is indeed true for optically thin clouds, with total thickness less than 1, and for nonsaturated clouds. In case of AGNs, either observed in optical/UV or X-ray range, very often lines are saturated, and the gas with optical depth larger than unity is needed to explain all absorption features.   
The first brake to the degeneracy of photoionization models due to gas density was presented by 
\citet{rozanska2008} in the case of warm absorbers in AGN. In addition, it was broadly discussed
by \citet{adhikari2019}, where the direct physical reason for such behavior was provided. 

Today, we accept that for volume densities higher than $ \sim 10^{6-8}$\,cm$^{-3}$, depending on the shape of the illuminating broadband continuum, outgoing radiation depends not only on the ionization parameter and column density, but also on the volume density of the gas.
Since our model depends on many parameters, it was not possible to fit directly the volume density of the gas at the cloud surface. We clearly demonstrated that AMD for Mrk~509  can be well reproduced for a gas density of about $10^8$\,cm$^{-3}$ \citep{adhikari2015}. 

For the goal of this paper we computed models for various values of the gas density at the cloud surface and microturbulence. Since those models were huge it imposed serious problems while using it in {\sc XSPEC} with Monte Carlo fitting method. Thus, we decided to omit direct fitting of $n_{\rm H,0}$ and turbulence. We only compared best-fit solutions for log $n_{\rm H,0}$: 8, 10 and 12. The best C-stat was achieved for a value of 10, although the difference was small. After fixing the gas density, we proceed in a similar way with turbulence choosing 500 km s$^{-1}$ as preferred.
Therefore, the gas density and turbulence are not fully fitted but weakly constrained. It is just accepted to be the best density for which other parameters, such as the ionization parameter and column density, give an acceptable fit. 

We obtained this best fit for $n_{\rm H,0} = 10^{10}$~cm$^{-3}$, which places our WA cloud at a distance of $0.02$ pc, using the fitted value of the ionization parameter and the source luminosity of Sect.~\ref{sec:source}.
This distance is a factor of 4 higher than our earlier finding
\citep{adhikari2015,adhikari2019}, but it is in agreement with the location of the broad line region
found in the reverberation mapping. We suggest here that such WA can be connected with a BLR gas
of lower temperature and can be responsible for additional continuum emission observed by RGS detectors.

\subsection{WA lines} 
\label{sec:walines}

To take a closer look at our analysis, we focus on different spectral bands, presenting the overall best-fitted model together with the data in the observed frame in Figs. from~\ref{fig:linespanela}  to~\ref{fig:linespanele}. Both RGS spectra were fitted simultaneously with parameters listed in Tab.~\ref{tab:stat}.   The prominent absorption lines resulting from several ions of \element{Mg}, \element{Ne}, \element{Fe}, \element{O}, \element{N}, and \element{C} are presented in the figures. All lines associated with WA are identified with black labels, whereas those associated with Galactic absorption are identified with green labels. Lines that appear in RGS\,1 and are not detected by RGS\,2 are probably associated with instrumental lines. 
The instrumental lines that are included in the response matrix of both gratings are listed in Table~\ref{tab:instrument_lines} in Appendix~\ref{app:lines}.

\begin{table*}
	\caption{Absorption lines due to the WA in Mrk~509 identified with CTP model. } 
	\label{tab:lines}

\begin{tabular}{lrrllcl}
\hline\noalign{\smallskip}
 Ion & $\lambda$~~ & E~~ & EW & EW   & $N_{\rm ion}$ from WA model & Ion ~$\lambda$ [$\AA$] \\ 
 {\sc titan}  & [$\AA$] & [eV] & $\times 10^{-2}$ [$\AA$] & [eV] &  [cm$^{-2}$] & D11  \\ 
\noalign{\smallskip}\hline\noalign{\smallskip}
\ion{Mg}{XI}     &  9.49 &  1307.12  &   0.1740$_{-0.014}^{+0.015}$ & 0.256$_{-0.021}^{+0.022}$           &   3.37$_{-0.29}^{+0.31}  \, 10^{15}$  & \ion{Mg}{XI} ~ 9.46 \\
\ion{Ne}{IX}     & 11.95 &  1037.77  &  0.3463 $\pm$ 0.096   & 0.3218 $\pm$ 0.0089                        &   2.18$_{-0.06}^{+0.07}  \, 10^{16}$  & \ion{Ne}{IX}  ~11.92\\
\ion{Ne}{X}     & 12.57 &  986.36  &   0.266 $\pm$ 0.022 & 0.224 $\pm$ 0.019                             &  5.11$_{-0.44}^{+0.47}  \, 10^{15}$   & \ion{Ne}{X} ~12.54  \\
\ion{Fe}{XVII}   & 12.64 &  980.87  &   0.00269 $\pm$ 0.00057 & 0.00223 $\pm$ 0.00047                     &  2.74$_{-0.56}^{+0.59}  \, 10^{13}$   & \ion{Fe}{XXI} ~12.66 \\
\ion{Ne}{IX}     & 13.91 &  891.17  &  1.699 $\pm$ 0.033 &  1.164$_{-0.022}^{+0.023}$                     &  2.18$_{-0.06}^{+0.07}  \, 10^{16}$   & \ion{Ne}{IX} ~13.90  \\
\ion{Ne}{IX}     & 13.91 &  891.17  &   1.699 $\pm$ 0.033 &  1.164$_{-0.022}^{+0.023}$                    &  2.18$_{-0.06}^{+0.07}  \, 10^{16}$   & \ion{Fe}{XIX} ~13.98\\
\ion{Ne}{VIIIu}  & 14.12 &  878.37  &   0.202$_{-0.017}^{+0.018}$ &  0.135$_{-0.011}^{+0.012}$            &  3.30$_{-0.28}^{+0.30}  \, 10^{15}$   & \ion{Ne}{VIII} ~14.12 \\
\ion{Ne}{VIIu}   & 14.29 &  867.69  &   0.207$_{-0.030}^{+0.029}$ &  0.135 $\pm$ 0.019                    &  2.04$_{-0.27}^{+0.30}  \, 10^{15}$   & \ion{Ne}{VII} ~14.28 \\
\ion{Fe}{XVu}  & 14.68 & 844.82 & 0.00262$_{-0.00013}^{+0.00012}$ & 0.001607$_{-0.000080}^{+0.000069}$    &  7.88$_{-0.84}^{+0.94}  \, 10^{13}$   & \ion{Fe}{XVIII} ~14.68 \\
\ion{O}{VIII}    &   15.34 &  808.46  &   0.260 $\pm$ 0.011 &  0.1468 $\pm$ 0.0062                        &  9.83$_{-0.42}^{+0.44}  \, 10^{16}$   & \ion{O}{VIII} ~15.33 \\
\ion{O}{VIII}    &   15.70 &  789.50  &   0.529 $\pm$ 0.021 &  0.284$_{-0.011}^{+0.012}$                  &  9.83$_{-0.42}^{+0.44}  \, 10^{16}$   & \ion{O}{VIII} ~15.69\\
\ion{O}{VIII}    &   16.56 &  748.58  &   1.390 $\pm$ 0.047 &  0.672$_{-0.022}^{+0.023}$                  &  9.83$_{-0.42}^{+0.44}  \, 10^{16}$   & \ion{O}{VIII} ~16.55\\
\ion{O}{VIIs}    &   18.03 &  687.68  &   0.5670 $\pm$ 0.039 &  0.232 $\pm$ 0.016                         &  8.52$_{-0.61}^{+0.65}  \, 10^{16}$   & \ion{O}{VII} ~18.02 \\
\ion{O}{VII}     &   18.38 &  674.52  &   1.0840 $\pm$ 0.069 &  0.426 $\pm$ 0.027                         &  8.52$_{-0.61}^{+0.65}  \, 10^{16}$   & \ion{O}{VII} ~18.38 \\
\ion{O}{VII}     &   19.27 &  643.38  &   2.44$_{-0.11}^{+0.12}$ & 0.871 $\pm$ 0.041                      &  8.52$_{-0.61}^{+0.65}  \, 10^{16}$   & \ion{O}{VII} ~19.27 \\
\ion{O}{VIII}    &   19.63 &  631.62  &   4.3680 $\pm$ 0.055 &  1.504$_{-0.019}^{+0.020}$                 &  9.83$_{-0.42}^{+0.44}  \, 10^{16}$   & \ion{O}{VIII} ~19.61 \\
\ion{S}{XIIs}    &   22.83 &  543.10  & 0.000588 $\pm$ 0.000059 &  0.000150 $\pm$ 0.000015                &  5.98$_{-0.57}^{+0.61}  \, 10^{14}$   & \ion{O}{VI} ~22.77 \\
\ion{S}{XIIIs}   &   23.08 &  537.26  & 0.00264 $\pm$ 0.00035 &  0.000657 $\pm$ 0.000087                  &  9.18$_{-1.10}^{+1.30}  \, 10^{13}$   & \ion{O}{V} ~ 23.12 \\
\ion{N}{VII}     &   25.64 &  483.51  &   2.312 $\pm$ 0.059 &     0.467$_{-0.011}^{+0.012}$               &  1.36$_{-0.04}^{+0.05}  \, 10^{16}$   & \ion{N}{VII} ~25.62 \\
\ion{C}{VI}      &   27.93 &  443.94  &   0.591 $\pm$ 0.039 &    0.1006 $\pm$ 0.0066                      &  3.32$_{-0.22}^{+0.24}  \, 10^{16}$   & \ion{C}{VI} ~27.93 \\
\ion{C}{VI}      &   29.45 &  420.94  &   1.568 $\pm$ 0.094 &     0.240$_{-0.014}^{+0.015}$               &  3.32$_{-0.22}^{+0.24}  \, 10^{16}$   & \ion{C}{VI} ~29.44 \\
\ion{N}{VI}      &   29.78 &  416.33  &    1.74$_{-0.16}^{+0.17}$ &     0.260 $\pm$ 0.025                 &  4.15$_{-0.44}^{+0.47}  \, 10^{15}$   & \ion{N}{VI} ~29.77 \\
\ion{S}{Xs}      &   30.50 &  406.45  &  0.0559$_{-0.0013}^{+0.0014}$ &   0.00797$_{-0.00019}^{+0.00020}$ &  1.30$_{-0.03}^{+0.03}  \, 10^{15}$   & \ion{N}{V} ~30.40 \\
\ion{C}{VI}      &   34.91 &  355.15  &  6.39$_{-0.19}^{+0.20}$ & 0.695$_{-0.021}^{+0.022}$               &  3.32$_{-0.22}^{+0.24}  \, 10^{16}$   & \ion{C}{VI} ~34.89 \\
\ion{C}{V}       &   36.17 &  342.76  &  0.446$_{-0.063}^{+0.061}$ & 0.0452$_{-0.0064}^{+0.0062}$         &  3.01$_{-0.41}^{+0.44}  \, 10^{15}$   & \ion{C}{V} ~36.17 \\
\noalign{\smallskip}\hline
\end{tabular}

\tablefoot{
The first, second, and third columns give the line name, its observed wavelength, and energy in the {\sc titan} models. The fourth, fifth, and sixth columns provide the equivalent widths of the given lines in the model components and their corresponding ionic column densities. The last column refers to the line name and observed wavelength, following \cite{kaastra2011}.
The EW and $N_{ion}$ values together with their uncertainties are interpolated from the nearest grid points to correspond to the solution of $\log \xi_0 = 1.9828_{-0.0089}^{+0.0093}$, $\log N_{\rm H} = 20.456 \pm 0.016$.
}
\end{table*}

The identification of absorption lines due to the WA medium in AGN is associated on the basis of known atomic data of X-ray transitions. 
There has been a long-standing discussion about a common database of atomic transitions, but to date it does not exist, and the data may differ slightly in various theoretical models. 
In general, our identification of lines is almost the same as in \citetalias{detmers2011}, with small exceptions of several transitions, which are located very close to each other in energy scale. Such blended lines are difficult to disentangle with the current resolution of X-ray data. All lines associated with WA and identified by us are listed in Table~\ref{tab:lines}. 

\begin{figure}
  \centering
  \includegraphics[width=0.99\linewidth]{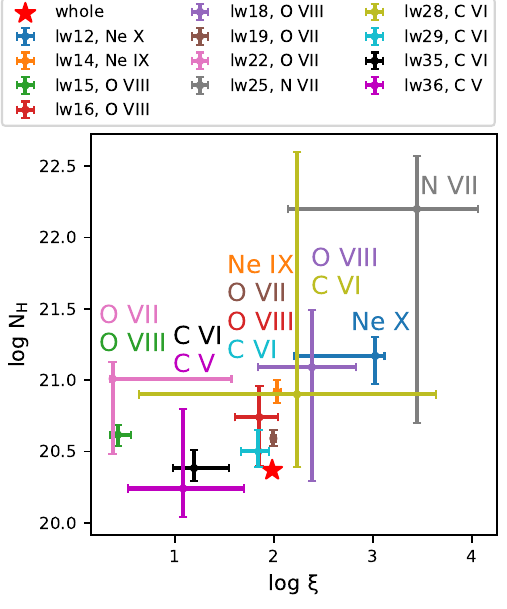}
  \caption{Best fitted parameters using spectral windows with warm500 multiplicative model.}
 \label{fig:lwfits}%
\end{figure}

To find the optimal parameters of the model and reproduce prominent lines, a test was prepared. The spectra were divided into windows that contained prominent absorption lines. The model of log $n_{\rm H,0}$ = 10, $v_{\rm turb}$ = 500 km s$^{-1}$, solar abundance was used (warm500). Spectral windows were named lwxx, where xx denotes the integer wavelength value contained in the window. Most of the fits in these spectral windows achieve a c-stat / dof in the range $1.2-1.8$. 
The fitting parameters are also illustrated in Fig.~\ref{fig:lwfits}. We see that the best fit of the entire spectrum with a single absorber marked with a red star is located in the middle of log $\xi$. However, the total column density is low to avoid stronger absorption that is not visible in the spectrum and preserve a good fit of the continuum. As a consequence of the low total column density, the lines are non-saturated, therefore both CTP and CD models may successfully reproduce the data.

We calculated the velocity shift of the WA lines (average value of the prominent absorption lines) to be $v$=-197 [km~s$^{-1}$] (blue shift/outflow) in the rest-frame.  
These values are presented in Tab.~\ref{tab:stat}. 
We adopt the hydrogen number density $n_{\rm H,0}$=$10^{10}$ cm$^{-3}$.
Moreover, given our low column density in the {\sc titan} calculations and with a high gas density, we never go through a general stratification of the ionization parameter. This is one of the reasons why we do not have a full s-curve as seen from the middle panel of Fig.~\ref{fig:ionization}.

\section{Discussion}
\label{sec:disc}

We performed analysis of high-resolution RGS XMM-Newton data from \object{Mrk 509} with a model of a stratified medium in the constant total pressure (CTP) regime. Our data reduction procedure allowed us to collect 900 ks observations - the longest possible for this source. Both RGSs have been fitted simultaneously, and we have taken care of proper data binning without loss of spectral resolution.

Our results are comparable to those obtained during the most detailed fit of the absorption spectra of \object{Mrk 509} published by \citetalias{detmers2011}. The authors fine-tuned the input ionizing continuum and describe it with splines to achieve a better match in the spectra. 
Here we use the same SED as \citetalias{detmers2011} following the work of \citet{kaastra2011a}.
However, we incorporate a slightly different data set, RGS observations collected over the longer period. It is potentially the source of the short-wavelength part difference.

\citetalias{detmers2011} conducted a very detailed fitting of the absorption lines in the \object{Mrk 509} spectrum. This allowed reconstructing smooth AMD made up of many layers responsible for the effective creation of different absorption lines. However, they concluded that the absorption spectrum can be reasonably and accurately approximated with the five constant density slabs model. These components are spread over the scale of ionization parameters; 
log$\xi$: -0.14, 0.81, 2.03, 2.2, 2.62, 3.26 for their model 3 (their table 6).
The corresponding values of the column density of the slabs are: 19.6, 19.9, 20.41, 20.64, 20.25, 20.8 in logarithmic cm$^{-2}$.
While \cite{ebrero2011} found log $\xi$: 1.06, 2.26, 3.15 with log column densities: 20.3, 20.7, 20.8 by fitting soft X-ray spectrum taken with Chandra.

Our best-fit static CTP model with the fitted parameters: $\log \xi = 1.9829 ^{+0.0093}_{-0.0089}$ and $\log N_{\rm H} = 20.456 \pm 0.016$ for WA outflow with $v=195\pm16$~km~s$^{-1}$ 
is slightly worse than that of \citetalias{detmers2011}, but is satisfactory for most of the lines, and the overall fit looks good with C-stat/dof = 1.9.
 Nevertheless, we have checked that for our model and these particular data sets, increasing CTP model components (i.e. increasing $i$) does not improve our fit considerably, only by C-stat of a few. Therefore, we refrain from adding additional WA model components,
because it only complicates our analysis, and therefore only a single CTP model is discussed here. 
However, the single model was dominant in most of the absorption lines except the \ion{Ne}{X} line.
One thing is the velocity gradient between different absorption lines, which limits fit quality of a single static photoionization model. The velocity shift issue could be resolved by adding more shifted components between each other, but the expected improvement was small. The ultimate difficulty was that with the assumed ionizing continuum as \citetalias{detmers2011}, improving all prominent lines at once causes the emergence of additional absorption, which destroyed the continuum fit in the absorption line-free area. This was caused by the need to increase the column density in a larger $\xi$ space (addressing \ion{Ne}{X}, as in the solution 'lw12'). It should be noted that we do not address a cooler medium below 10$^{4}$ K in our model. 

Our solution, however, not as good as Detmer's, is close in ionization parameters and column densities. Our S-curve is also similar. The main difference is that the solutions computed in {\sc titan} reside in $\Xi$ lower than the solutions proposed by \citetalias{detmers2011} for the corresponding values of $\xi$. $\Xi$ is the dynamical ionization parameter and is defined as $\Xi \equiv P_{rad}/P_{gas}$, where $P_{rad}$ is the radiation pressure and $P_{gas}$ is the gas pressure. The difference is around -0.2 in log $\Xi$ for models with log $n_{{\rm H,0}}$ = 8 on the face of the cloud. The difference is greater, around -0.4, for log $n_{\rm H,0}$ = 12.
However, there are a few differences in our approach, such as the use of a different ionization code, slightly different abundances and atomic data, a data set covering a broader time span without binning, and fitting as two detector spectra at once.

To address the issue of a few unfitted lines and recheck those that were fitted, we repeated the fitting using narrower windows containing individual strong absorption lines.
The results of this procedure are 
presented in Fig.~\ref{fig:lwfits}. It shows that, in fact, the optimal ionization parameters might be spread over a range of values from ~0.5 to ~3.5 in $\log \xi$. Although huge error bars are present for some lines indicating that a broad range of $\xi$ is possible (such as for \ion{N}{VII} in lw28). The problem here is that the windows are narrow and do not address the impact of the absorbing cloud on the whole continuum. For several lines, the "optimal" column density is large, and affected continuum matches outside of a given window. This is the reason why the overall fitting does not reach higher column densities. However, the results in Fig.~\ref{fig:lwfits} seem to be consistent with the solution of \citetalias{detmers2011} and \cite{ebrero2011}.

We can place our WA cloud at a distance of 0.02 pc or $6 \times 10^{16}$ cm, using the fit value of the ionization parameter, $n_{\rm H,0} = 10^{10}$ cm$^{-3}$  and the source luminosity. This value indicates the overdensity in the atmosphere of the accretion disk and the thickening of the disk \citep{adhikari2018ilr}. That would be close to the inner broad line region position. However, it is much smaller than the limits proposed by \cite{ebrero2011, kaastra2012}, who estimated from the X-ray spectra modeling that the absorber should be farther than 5 pc, because of its size-radius constraint. However, based on $\xi$ variability \cite{kaastra2012} suggested an upper limit of 0.5 pc, which would be consistent with our estimate.

\subsection{\ion{Ne}{X} problem}
\begin{figure}
  \centering
  \includegraphics[width=0.99\linewidth]{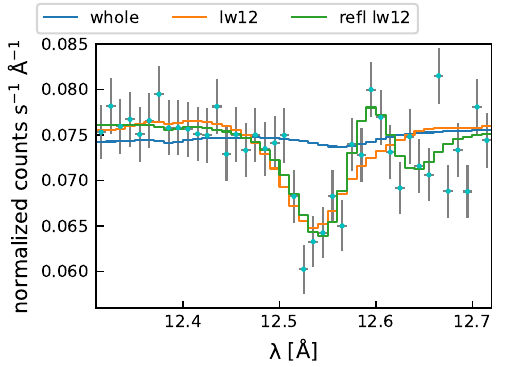}
  \caption{Spectrum in the wavelength range of 12.3--12.7\ \AA\ containing \ion{Ne}{X} line.}
 \label{fig:lw12fit}%
\end{figure}

This global solution is not able to reproduce Ne lines --- especially \ion{Ne}{X}, which is optimally fitted with a higher ionization parameter and column density. 
Fitting in the window spanning from 12.3 up to 12.7 \AA\ (model "lw12") gives good results.
\ion{Ne}{X} $\lambda \sim$ 12.5 \AA\ is possible to fit for a model with  $\log \xi > 1.8$ and $\log N_{\rm H} > 21.1$. The exact values depend on the configuration of the additional components (such as reflection). The region of the \ion{Ne}{X} line is plotted in Fig.~\ref{fig:lw12fit}, where we see two solutions that reproduce the line: orange and green, while the best global solution is plotted in blue. Orange is for the multiplicative model with the incident continuum fitted in lw12 window, which gives: log $\xi$ = 2.79, log $N_{\rm H}$ = 21.17; green shows the additive model with reflection fitted in lw12 window, where the absorption and reflection parameters are: log $\xi$ = $1.82^{+0.68}_{-1.1}$, log $N_{\rm H}$ = $21.45^{+0.36}_{-0.24}$.
However, what these two models have in common, in contrast to the global solution, is primarily a higher column density and a blueshift of the order of 1000 km s$^{-1}$. Meanwhile, fitting other lines and/or reflection requires lower shifts.

\cite{ebrero2011} concluded that their low $ \log \xi \sim 1$ component is not in pressure equilibrium with the other two, which are more ionized. In principle, they did not directly reject the possibility of the coexistence of the $\log \xi = 2.3$ and $\log \xi = 3.2$ solutions being in pressure equilibrium.
Our single cloud solution gives a good overall fit and is consistent with the $\log\xi \sim 2$ medium.
However, the problem with reproducing \ion{Ne}{X} absorption and the fact that it is more shifted than the oxygen lines indicate that a more complex model is needed. A second component or a dynamic model that accounts for medium expansion is likely required.

It was tempting to test whether a continuous AMD \citep{adhikari2015} could be reproduced by directly fitting the spectrum with a single CTP model. However, this was not possible with direct fitting of a single static CTP model. One of the key issues was the gradient in the line shifts, which suggests that dynamics must be addressed in physically motivated models.

\subsection{Emission}
\citetalias{detmers2011} identified emission features as radiative recombination continua, the most prominent around 22--24~\AA\ from the \ion{O}{VII} lines. Our trials with using the reflection spectrum over the same TITAN grid as absorption one did not improve the fit. The sharp emission lines present in the model appeared in a broad range of wavelengths. It was necessary to broaden the emission spectrum. The flux excess could potentially be addressed only by a smeared reflection. However, our attempt to apply a smeared reflection component also does not improve the fit globally. Since our reflection spectra show emission lines in a broader range of wavelengths, optimizing the fit solution resulted in negligible emission model amplitude. However, locally emission indeed improved the fit, as in the \ion{Ne}{X} window presented in Fig.~\ref{fig:lw12fit}. It is possible that 
the illuminating continuum that is used in our case may not represent the total spectrum seen by the absorber.
Since implemented, the continuum is a spline that incorporates bumps into the continuum. It is possible that as a result the possible flux excess coming from a smeared emission is incorporated into the assumed continuum. Fitting reflection from TITAN would require revisiting the ionizing continuum and recomputing the grid.
This is beyond the scope of this work.

\section{Conclusions}
\label{sec:conc}

The CTP model with $\log \xi = 1.9829^{+0.0093}_{-0.0089}$ and $\log N_{\rm H} = 20.456 \pm 0.016$ for the WA outflow with velocity $v=195\pm16$~km~s$^{-1}$, as presented in our work, provides a good general fit to the XMM RGS data. We found that a static CTP model does not address all WA absorption lines with a single component. This is due to the fact that it does not include velocity gradient across the slab and thus does not fit gradual line shift with the ionization parameter. 
Although preserving the apparent continuum shape in the fit does not allow us to increase the column density above $\log N_{\rm H} \sim 20.5$, an additional component with $\log \xi \sim 3$ is required to account for \ion{Ne}{X} absorption, which is consistent with the conclusions in the literature \citep[\citetalias{detmers2011}]{ebrero2011}.

We conclude that the CTP model is applicable, while more self-consistent than CD models, and is more straightforward to interpret. We were unable to reproduce a continuous AMD by direct fit of the single static CTP model, most probably because of the dynamics of the observed medium.

\begin{acknowledgements}
    We are grateful to the anonymous referee for his/her comments that improved our paper. 
    The authors thank Alex Markowitz for helpful discussion. This research was partially supported by the Polish National Science Center grant No. 2021/41/B/ST9/04110. TPA acknowledges the support from the National Natural Science Foundation of China (nos. 12222304, 12192220, and 12192221).

\end{acknowledgements}

\bibliographystyle{aa} 
\bibliography{models}

\begin{appendix}

    \begin{figure*}
   \centering
   \includegraphics[width=0.99\linewidth]{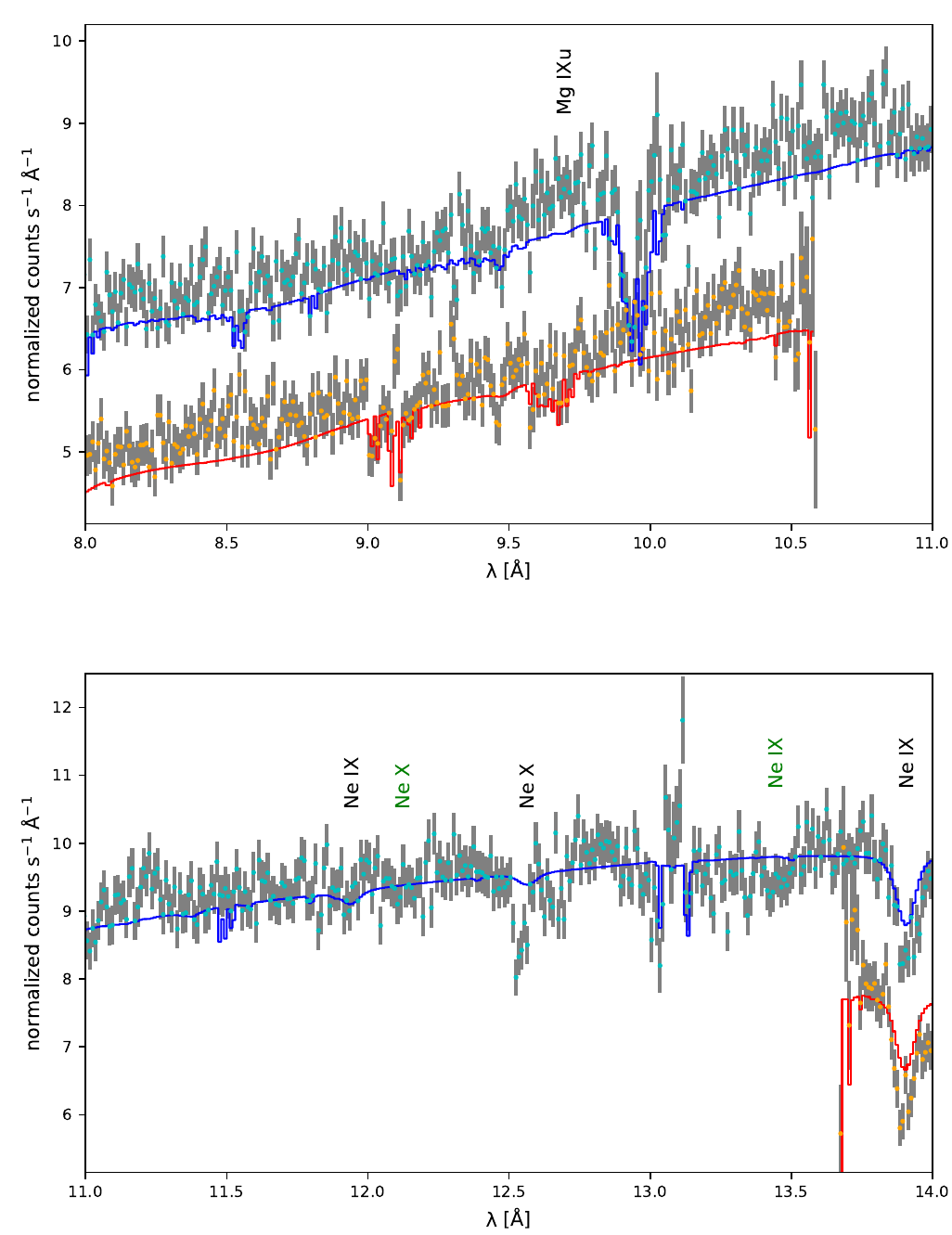}
   \caption{Overall model fit to \object{Mrk 509} RGS spectrum given in observed frame. RGS\,1 and RGS\,2 spectra are plotted in orange and cyan dots with arbitrary shift 0.02 counts s$^{-1}$ \AA$^{-1}$, and  errors marked by gray bars.  The total model: continuum together with single CTP WA is plotted in red and blue. 
   WA lines are identified with black text, while possible absorption lines on Milky Way - by green text. For instrumental lines see Appendix~\ref{app:lines}.}
              \label{fig:linespanela}
    \end{figure*}
  \begin{figure*}
   \centering
   \includegraphics[width=0.99\linewidth]{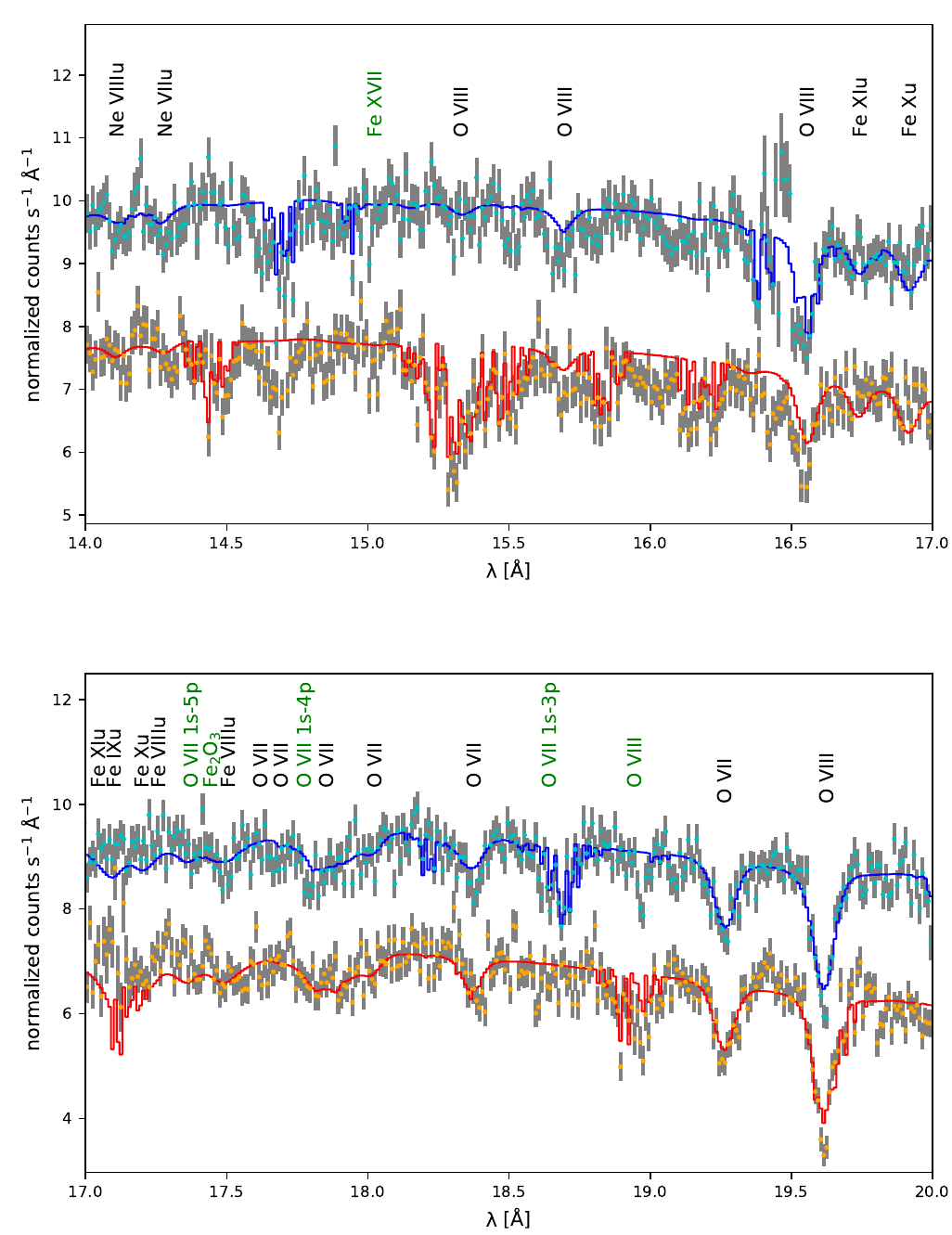}
   \caption{continued.}
              \label{fig:linespanelb}
    \end{figure*}
  \begin{figure*}
   \centering
   \includegraphics[width=0.99\linewidth]{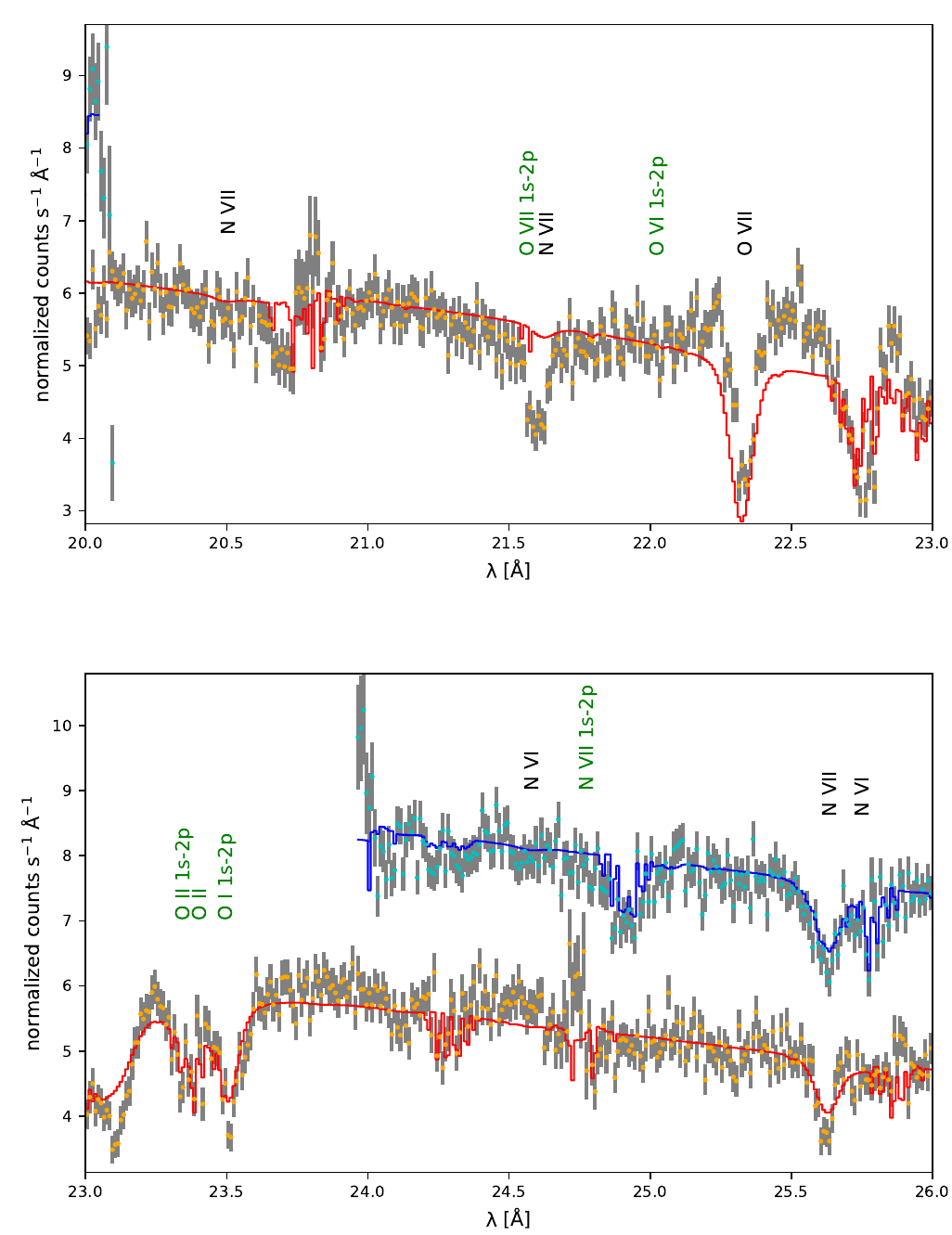}
   \caption{continued.}
              \label{fig:linespanelc}
    \end{figure*}
  \begin{figure*}
  \centering
  \includegraphics[width=0.99\linewidth]{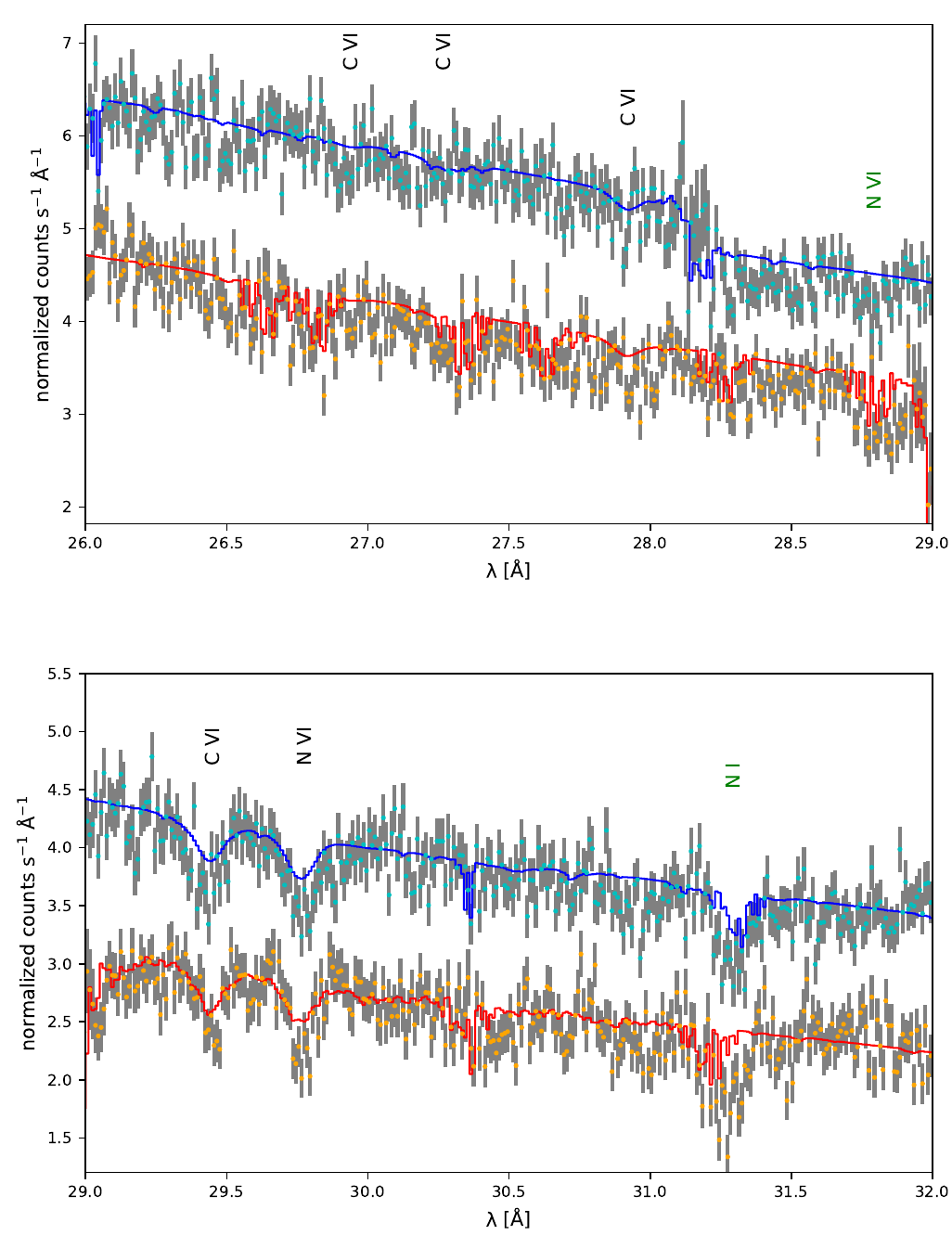}
  \caption{continued.}
            \label{fig:linespaneld}
  \end{figure*}
  \begin{figure*}
  \centering
  \includegraphics[width=0.99\linewidth]{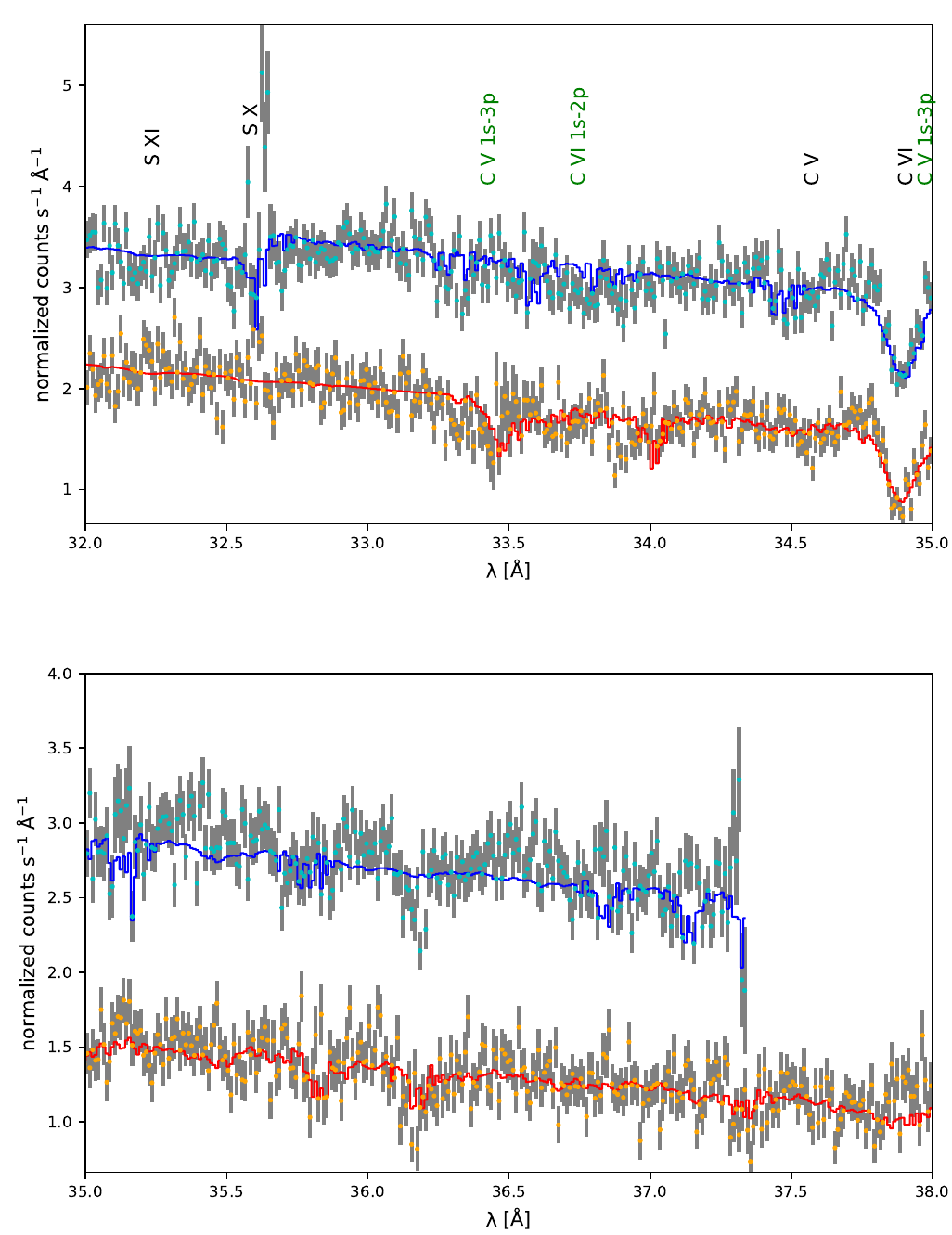}
  \caption{continued.}
              \label{fig:linespanele}
  \end{figure*}

\section{Absorption lines of an external origin} 
\label{app:lines}

External absorption originating in the Milky Way may display the same lines as those of WA, which may provide additional complication in the analysis of the intrinsic absorption by the ionized gas at the source rest frame.
In order to determine which lines originate externally to the source, we perform an exercise by fitting multiphase Galactic absorption with a more advanced model.
The term responsible for the Galactic absorption in the cold phase was fitted by the {\sc ismabs} 
model \citep{gatuzz2015}, while the warm and hot Galactic phases \citep{nicastro2016, gatuzz2016} were modeled through the models
{\sc ismabs} and {\sc ioneq} \citep{gatuzz2018}, respectively. 
The fitted values of the parameters for the continuum
model responsible for Galactic absorption are given
in Tabs.~\ref{tab:ismabs} and \ref{tab:ioneq}. The column density of neutral hydrogen for the cold/warm
phase in the {\sc ismabs} model was fixed at $4.44 \times 10^{20}$ cm$^{-2}$ \citep{murphy1996}.
Tab.~\ref{tab:ismabs} displays all three ion populations of this model, i.e. neutral, single, and double ionized species, but only detectable
ions are listed.   In the hot phase fitted by the {\sc ioneq} model, we left the column
density of the neutral hydrogen as a free parameter. Only detectable ions are listed
in Tab.~\ref{tab:ioneq}. 
We also tested a three-component {\sc ioneq} model instead of the aforementioned combination of {\sc ismabs} and a single {\sc ioneq}, but this approach did not result in better fit quality.

\begin{table*}
\caption{{\sc ismabs} model parameters fitting cold Galactic absorption.
}
\label{tab:ismabs}
\begin{center}
\begin{tabular}{lrrrrrr}
\hline\noalign{\smallskip}
type of ion's population & H  [$\times 10^{20}$] &  C  [$\times 10^{16}$]      & N    [$\times 10^{16}$]   & O  [$\times 10^{16}$]    & Ne   [$\times 10^{16}$]    & Mg   [$\times 10^{16}$]   \\

\noalign{\smallskip}\hline\noalign{\smallskip}
     & 4.44* &     &       &       &       & \\
I   &       & 2 $\pm 210$ & 4.24 $\pm 0.90$ & 20.9 $\pm 1.2$ & 2.4
$\pm 2.5$ & 1 $\pm 32$  \\
II  &       & 0.0 $\pm 1.5$ & 0.30 $\pm 0.13$ & 0.01 $\pm
0.27$ & 9.4 $\pm 1.5$ & 0 $\pm 58$  \\
III &        & 0 $\pm 150$ & 0.045 $\pm 0.076$ & 1.65 $\pm 0.32$ &
2.3e-6 $\pm 1$ & 8 $\pm 30$   \\
\noalign{\smallskip}\hline
\end{tabular}
\end{center}
\tablefoot{
Column densities for a certain ion are given in the units of cm$^{-2}$.
Missing ions were not detectable and are not listed in the table. Neutral hydrogen
column density for {\sc ismabs} marked by star was taken from the literature and used as a
fixed value.
}
\end{table*}

\begin{table*}
\caption{{\sc ioneq} model parameters fitting warm/hot Galactic absorption.  }
\label{tab:ioneq}
\begin{center}
\begin{tabular}{lrrrrrrr}
\hline\noalign{\smallskip}
$\log T$ [K]  & $\log \xi$ [erg~cm~s$^{-1}$] & H [$\times 10^{20}$]  & O [$\times 10^{16}$]   &  Fe [$\times 10^{16}$] & $v_{\rm turb}$ [km~s$^{-1}$] \\

\noalign{\smallskip}\hline\noalign{\smallskip}
  5.579 $\pm 0.012$  & 6.84 $\pm 0.15$ & 3.42 $\pm 0.61$  & 1.39 $\pm
0.26$     & 2.17 $\pm 0.85$ & 14.3 $\pm 2.1$  \\
\noalign{\smallskip}\hline
\end{tabular}
\end{center}
\tablefoot{
Mostly hot phase containing iron was fitted. Column
densities for a certain element are given in the units of cm$^{-2}$.
}
\end{table*}

By fitting {\sc ioneq} and {\sc ismabs} models, we were able to identify several Galactic absorption lines in our data, which are taken into account while identifying warm absorber lines in Sect.~\ref{sec:walines}. 
All Galactic absorption lines that may influence our analysis were cross-checked with the
papers by \citet{kaastra2009,kaastra2011,detmers2011,kaastra2018}, and are listed in the Tab.~\ref{tab:ism_lines}.
These lines are marked with green labels in Figs.~\ref{fig:linespanela}  to~\ref{fig:linespanele} that present the final fit of the CTP WA model.

\begin{table}
\caption{List of Galactic absorption lines identified during {\sc ismabs} and {\sc ioneq} model fitting to the Mrk509 RGS\,1 and RGS\,2 data. }
\label{tab:ism_lines}
\begin{center}
\begin{tabular}{lcc}
\hline\noalign{\smallskip}
Label & $\lambda$ [\AA] & reference\\
\noalign{\smallskip}\hline\noalign{\smallskip}
\ion{Ne}{X} & 12.10--12.15 & 1\\
\ion{Ne}{IX} &13.454 & 2\\
\ion{Fe}{XVII}& 15.00--15.05 & 1\\
\ion{Fe}{IX} & 16.50--16.60 & 1\\
\ion{O}{VII} (1s-5p)& 17.35--17.40 & 1\\
\ion{O}{VII} (1s-4p)& 17.75--17.80 & 1\\
\ion{O}{VII} (1s-3p)& 18.624 & 2\\
\ion{O}{VII} (1s-3p) & 18.60--18.70 & 1\\
\ion{O}{VIII}        & 18.90--19.00 & 1\\
\ion{O}{VIII} (1s-2p) & 18.966 & 2\\
\ion{O}{VII} (1s-2p) & 21.55--21.65 & 1\\
\ion{O}{VII} (1s-2p) &21.604 & 2\\
\ion{O}{VI} (1s-2p)  &22.025 & 2\\
\ion{O}{VI} (1s-2p)  & 22.00--22.05 & 1\\ 
\ion{O}{II} (1s-2p)  & 23.30--23.40 & 1\\
\ion{O}{I} (1s-2p)   & 23.45--23.55 & 1\\
\ion{O}{I} (1s-2p)   & 23.521  & 2\\
\ion{N}{VII} (1s-2p) &24.75--24.80 & 1\\
\ion{N}{VI}  & 28.8  & 3\\
\ion{N}{I} (1s-2p) & 31.302 & 2\\ 
\ion{C}{V} (1s-3p) & 33.43 & 1\\
\ion{C}{VI} (1s-2p) & 33.743 & 2\\
\ion{C}{VI} (1s-2p) & 33.70--33.80 & 1\\
\ion{C}{V} (1s-3p)  & 34.95--35.00 & 1\\
\noalign{\smallskip}\hline
\end{tabular}
\end{center}
\tablefoot{
The first column shows the ion and its transition level, the second column describes the line wavelength in the rest frame and the references for the line identification are shown in the last column.
}
\tablebib{
(1)~\citet{kaastra2018}; (2) \citet{kaastra2011}; (3) \citet{detmers2011}.
}
\end{table}

In addition, care is taken to identify instrumental features that are potentially not included in the response matrix of the RGS instrument. The instrumental lines together with the references are listed in Table ~\ref{tab:instrument_lines}. For clarity of the figures presented in the final fit of high-resolution RGS data, we did not mark instrumental lines in the main body of the paper.

\begin{table}
\caption{List of the instrumental lines not included in the response matrix of RGS.
}
\label{tab:instrument_lines}
\begin{center}
\begin{tabular}{lccc}
\hline\noalign{\smallskip}
Label & $\lambda$ [\AA]  & reference\\
\noalign{\smallskip}\hline\noalign{\smallskip}
feature & 23.0 & 1\\
N$_{2}$\,H$_{4}$ & 30.6 & 1\\
hot pixel & 31.1 & 1\\
transition & 30.49 & 1\\
transition & 30.97 & 1\\
residua & 30.8 & 1\\
residua & 31.2 & 1\\
residua & 23.4 & 1\\
Mg\,F$_{2}$ (Mg edge) & 9.5 & 2\\
Mg\,F$_{2}$ (F edge) & 17.9 & 2\\
Al edge & 8.3 & 2\\
O absorption & 23 & 2\\
residua & 23.66 & 3\\
residua & 23.96 & 3\\
dip & 31.47 & 3\\
dip & 21.49 & 3\\
dip & 21.64 & 3\\
residua & 17.84 & 3\\
residua & 18.23 & 3\\
edge & 14.59 & 3\\
dip  & 13.2 & 3\\
\noalign{\smallskip}\hline
\end{tabular}
\end{center}
\tablefoot{
These lines are not marked in the plots for the clarity of the figures and are presented in this table instead.
}
\tablebib{
(1)~\citet{kaastra2018}; (2) \citet{C17}; (3) \citet{kaastra2009}.
}
\end{table}

\end{appendix}

\end{document}